# A single psychotomimetic dose of ketamine decreases thalamocortical spindles and delta oscillations in the sedated rat

Mahdavi A(1,2,3,4)*, Qin Y(1,2,3,5)*, Aubry A-S(1,2,3), Cornec D(1,2,3), Kulikova S(1,2,3,6), Pinault D(1,2,3)#.

(1) INSERM U1114, Neuropsychologie cognitive et physiopathologie de la schizophrénie, Strasbourg, France.
(2) Université de Strasbourg, Strasbourg, France
(3) Fédération de Médecine Translationnelle de Strasbourg, FMTS, Faculté de médecine, Strasbourg, France.
(4) University of Freiburg, Bernstein Center Freiburg, Germany
(5) Netherlands Institute for Neuroscience, Amsterdam, The Netherlands
(6) National Research University Higher School of Economics, Perm, Russia

(*) AM and YQ contributed equally to this work
(#) Corresponding author

Running title: Ketamine reduces sleep spindles

Number of words (references excluded): Whole ms: 5067; Abstract: 250; Introduction: 487; Methods and Materials: 916; Results: 1541; Discussion: 2123

Number of figures: 8

Number of tables: 0

Supplementary material: 11

Number of references: 129

Number of pages: 28

Correspondence to:

Didier PINAULT, PhD

INSERM U1114, Université de Strasbourg, FMTS

Faculté de Médecine

11, rue Humann

F-67085 Strasbourg Cedex, France

Tel: +33 (0)3 6885 3245

Email: pinault@unistra.fr

## ABSTRACT

BACKGROUND: In patients with psychotic disorders, sleep spindles are reduced, supporting the hypothesis that the thalamus and glutamate receptors play a crucial etio-pathophysiological role, whose underlying mechanisms remain unknown. We hypothesized that a reduced function of NMDA receptors is involved in the spindle deficit observed in schizophrenia.

METHODS: An electrophysiological multisite cell-to-network exploration was used to investigate, in pentobarbital-sedated rats, the effects of a single psychotomimetic dose of the NMDA glutamate receptor antagonist ketamine in the sensorimotor and associative/cognitive thalamocortical (TC) systems.

RESULTS: Under the control condition, spontaneously-occurring spindles (intra-frequency: 10-16 waves/s) and delta-frequency (1-4Hz) oscillations were recorded in the frontoparietal cortical EEG, in thalamic extracellular recordings, in dual juxtacellularly recorded GABAergic thalamic reticular nucleus (TRN) and glutamatergic TC neurons, and in intracellularly recorded TC neurons. The TRN cells rhythmically exhibited robust high-frequency bursts of action potentials (7 to 15 APs at 200-700Hz). A single administration of low-dose ketamine fleetingly reduced TC spindles and delta oscillations, amplified ongoing gamma-(30-80Hz) and higher-frequency oscillations, and switched the firing pattern of both TC and TRN neurons from a burst mode to a single AP mode. Furthermore, ketamine strengthened the gamma-frequency band TRN-TC connectivity. The antipsychotic clozapine consistently prevented the ketamine effects on spindles, delta- and gamma-/higher-frequency TC oscillations.

CONCLUSION: The present findings support the hypothesis that NMDA receptor hypofunction is involved in the reduction in sleep spindles and delta oscillations. The ketamine-induced swift conversion of ongoing TC-TRN activities may have involved at least both the ascending reticular activating system and the corticothalamic pathway.

**Abbreviations:**

AP-s, action potential-s

CT, corticothalamic

EEG, electroencephalogram

FFT, Fast Fourier Transformation

hfBurst, high-frequency burst of action potentials

NMDA, N-methyl d-aspartate

Non-REM, non-rapid eye movement

REM, rapid eye movement

TC, thalamocortical

TRN, thalamic reticular nucleus

## INTRODUCTION

Sleep abnormalities are detected not only during the early course of complex mental health diseases, such as schizophrenia (Kamath et al., 2015; Monti and Monti, 2005; Wamsley et al., 2012) but also in individuals having a high-risk mental state for developing a transition to psychotic and bipolar disorders (Zanini et al., 2015). Cortical EEG studies conducted in such patients have revealed a reduction in sleep spindles (Castelnovo et al., 2017; Ferrarelli et al., 2007; Ferrarelli et al., 2010; Manoach et al., 2014; Manoach et al., 2016) and slow-wave activity (Kaskie and Ferrarelli, 2018). The underlying neural mechanisms are unknown.

Sleep spindles have a thalamic origin with the GABAergic thalamic reticular nucleus (TRN) being a leading structure in their generation by exerting a powerful rhythmic inhibitory modulation of thalamocortical (TC) activities (Pinault, 2004; Steriade et al., 1985; Steriade et al., 1993). The TRN, the principal inhibitory structure of the dorsal thalamus, is innervated by two major glutamatergic inputs, TC and layer VI corticothalamic (CT) axon collaterals, which mediate most of their excitatory effects through the activation of glutamate receptors (Crandall et al., 2015; Deschênes and Hu, 1990; Gentet and Ulrich, 2003). Importantly, layer VI CT axons innervate simultaneously TC and TRN neurons (Bourassa et al., 1995), together forming a 3-neuron circuit robustly involved in the generation of sleep spindles (Bal et al., 2000; Bonjean et al., 2011). The specific, sensory and motor TC systems receive cortical inputs only from layer VI CT neurons, whereas the non-specific, associative/limbic/cognitive TC systems receive cortical inputs from both layer V and layer VI CT neurons (Guillery and Sherman, 2002). In contrast to layer VI CT neurons, layer V CT neurons do not innervate the TRN.

There is accumulating evidence that dysfunction of thalamus-related systems is a core pathophysiological hallmark for psychosis-related disorders (Andreasen, 1997; Clinton and Meador-Woodruff, 2004b; Cronenwett and Csernansky, 2010; Pinault, 2011; Steullet, 2019). NMDA receptors are also essential in the generation of thalamic spindles (Deleuze and Huguenard, 2016; Jacobsen et al., 2001), and a reduced function of these receptors is thought to play a critical role in the etio-pathophysiology of schizophrenia (Clinton and Meador-Woodruff, 2004a; Coyle, 2012; Krystal et al., 1994; Snyder and Gao, 2019; Vukadinovic, 2014). Furthermore, the NMDA receptor antagonist ketamine models a transition to a psychosis-relevant state in both healthy humans (Anticevic et al., 2015; Baran et al., 2019; Hoflich et al., 2015; Rivolta et al., 2015) and rodents (Chrobak et al., 2008; Ehrlichman et al., 2009; Hakami et al., 2009; Kocsis, 2012a; Pinault, 2008; Pitsikas et al., 2008). Therefore, we hypothesized that a reduced function of NMDA receptors is implicated in the reduction of the density of sleep spindles recorded in patients having or about to have psychotic disorders. In an attempt to test this hypothesis, the effects of a single low-dose of ketamine on sleep oscillations were investigated using network and cellular recordings in the dorsal thalamus and TRN along with an EEG of the frontoparietal cortex in the pentobarbital-sedated rat.

## METHODS AND MATERIALS

### Animals and drugs

Sixty-nine Wistar adult male rats (285-370 g) were used with procedures performed under the approval of the Ministère de l'Education Nationale, de l'Enseignement Supérieur et de la Recherche. Ketamine was provided from Merial (Lyon, France); clozapine, MK-801, apomorphine, and physostigmine, from Sigma-Aldrich (Saint-Quentin Fallavier, France), pentobarbital from Sanofi (Libourne, France), and Fentanyl from Janssen-CILAG (Issy-Les-Moulineaux, France).

### Surgery under general anesthesia

Deep general anesthesia was initiated with an intraperitoneal injection of pentobarbital (60 mg/kg). An additional dose (10-15 mg/kg) of pentobarbital was administered when necessary. Analgesia was achieved with a subcutaneous injection of fentanyl (10 $\mu$g/kg) every 30 min. The anesthesia depth was continuously monitored using an electrocardiogram, watching the rhythm and breathing, and measuring the withdrawal reflex. The rectal temperature was maintained at 36.5 °C (peroperative and protective hypothermia) using a thermoregulated pad. The trachea was cannulated and connected to a ventilator (50% air–50% $O_2$, 60 breaths/min). The anesthesia lasted about 2 h, the time necessary to perform the stereotaxic implantation of the electrodes (Pinault, 2005).

### Cortical EEG and thalamic cell-to-network recordings under sedation

To understand how ketamine could influence ongoing sleep oscillations, cortical EEG and cell-to-network recordings were performed in the TC system of pentobarbital sedated rats, a rodent model of slow-wave sleep with spindles (Connor et al., 2003; Ganes and Andersen, 1975; Pinault et al., 2006). At the end of the surgery, the rectal temperature was set to and maintained at 37.5°C. The analgesic pentobarbital-induced sedation was initiated about 2 h after the induction of the deep anesthesia and maintained by a continuous intravenous infusion of the following regimen (average quantity given per kg and per hour): Pentobarbital (4.2 ± 0.1 mg), fentanyl (2.4 ± 0.2 µg), and glucose (48.7 ± 1.2 mg). To help maintain a stable mechanical ventilation and to block muscle tone and tremors, a neuromuscular blocking agent was used (d-tubocurarine chloride: 0.64 ± 0.04 mg/kg/h). The cortical EEG and heart rate were under continuous monitoring to adjust the infusion rate to maintain the sedation.

For the cortical EEG recordings (28 rats), a recording silver wire (diameter: 200 $\mu$m) sheathed with Teflon was implanted in the parietal bone over the primary somatosensory cortex (from bregma: 2.3 mm posterior and 5 mm lateral).

The network and cellular recordings and labellings were done with glass micropipettes filled with a saline solution (potassium acetate, 0.5 M) and a neuronal tracer (Neurobiotin, 1.5%). Three series of experiments were carried out: 1) The first (16 rats) was designed to perform, along with the cortical EEG, extracellular (field potential and single/multiunit) recordings in specific and non-specific thalamic nuclei. The regions of interest were stereotaxically (Paxinos and Watson, 1998) located behind the bregma (2.3 to 3.6 mm posterior). 2) To consolidate the ketamine-induced effects on the extracellular recordings (population activities), a second series (8 rats) consisting paired juxtacellular TC and TRN recordings were performed in the somatosensory system. The diameter of the micropipette tip was about 1 $\mu$m (15-30 M$\Omega$) (Pinault, 1996). 3) In an attempt to understand the cellular membrane potential oscillations underlying the firing patterns, a third series of experiments (11 rats) was designed to record intracellularly TC neurons. The diameter of the micropipette tip was inferior to 1 $\mu$m (30–70 M$\Omega$). The extracellular and juxtacellular signals (0.1-6000 Hz), and the intracellular signal (0-6000 Hz) were acquired using a low-noise differential amplifier (DPA-2FL, npi electronic, GmbH) and an intracellular recording amplifier (NeuroData IR-283; Cygnus Technology Inc.), respectively. All signals were sampled at 20 kHz 16-bit (Digidata 1440A with pCLAMP10 Software, Molecular Devices). At the end of the recording session, the target neurons were individually labeled with Neurobiotin using the extra- or juxtacellular nano-iontophoresis technique (Pinault, 1996) to identify formally both the recording site and the structure of the recorded neurons (Fig1B). Then the animal was humanely killed with an intravenous overdose of pentobarbital, transcardially perfused with a fixative containing 4% paraformaldehyde in 10 mM phosphate buffer saline, and the brain tissue was processed using standard histological techniques for anatomical documentation.

## Data analysis

Analysis software packages Clampfit v10 (Molecular Devices) and SciWorks v10 (Datawave Technologies) were used. The spindle density was estimated by the number per 10 s of detected bouts filtered at the sigma-frequency (10-16 Hz) oscillations. Spectral analysis of EEG and network oscillations was performed with fast Fourier transformation (FFT, 2-Hz resolution). The power of baseline activity was analyzed in 4 frequency bands: delta-(1–4 Hz), sigma-(10–16 Hz, spindles), gamma-(30–80 Hz), and higher-(81–200 Hz) frequency oscillations. For each band, the total power was the sum of all FFT values. In single-unit juxtacellular recordings, single action potentials (APs) were detected using a voltage threshold and an inter-AP interval superior to 10 ms. High-frequency bursts (hfBursts) were identified based on a voltage threshold and an inter-AP interval inferior to 4 ms. A TC or TRN burst had a minimum of 1 inter-AP interval. Inter-AP time and autocorrelogram histograms, and the density (number per minute) of single APs and hfBursts were computed. To apprehend the time relationship between the network or cellular gamma waves and the cellular firing of a single TC or TRN neuron, a 25-55 Hz filter was used to make gamma waves detectable, to create a peri-event time histogram of the TC or TRN firings. Standard inter-AP interval (resolution 1 ms) histograms were computed. Each drug effect was measured relative to the vehicle condition with each rat being its control. Statistical significance of the observed effects was evaluated with the Student's paired t-test (significant when P≤0.05).

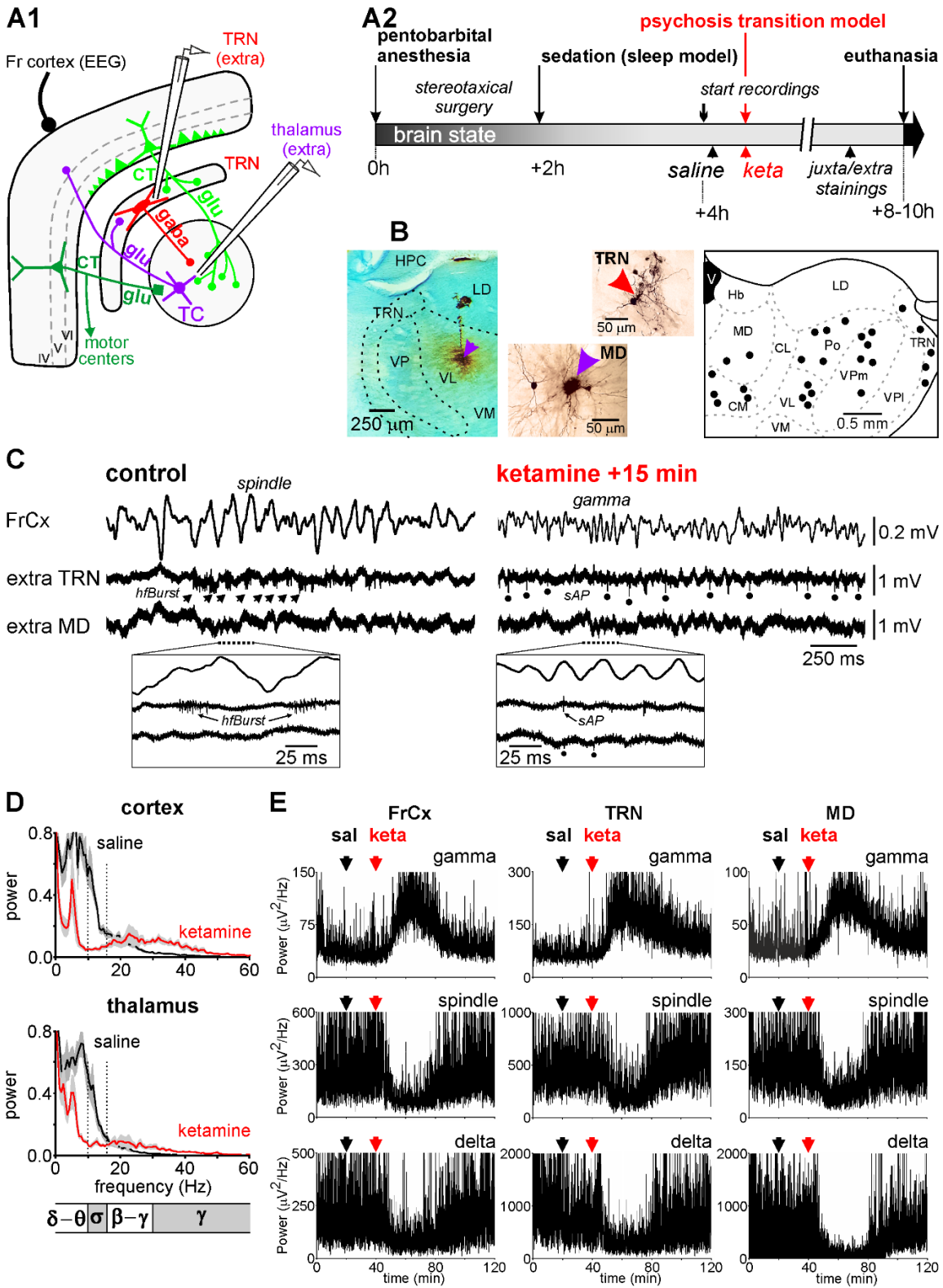

**Figure 1: Ketamine reduces sleep oscillations in the thalamocortical systems. (A1)** Experimental design showing the location of the two glass micropipettes designed to record the extracellular activities in the thalamic reticular nucleus (TRN) and in a dorsal thalamic nucleus along with the EEG of the frontal cortex. The hodology of the 4-neuron CT-TRN-TC circuit is also shown. The corticothalamic (CT) and thalamocortical (TC) neurons are glutamatergic while the TRN neuron is GABAergic. The cortical inputs of the first-order thalamic nuclei (like the ventral posterior, VP (somatosensory), and the ventral lateral, VL (motor)) originate from layer VI whereas the cortical inputs of the associative/limbic/cognitive thalamic nuclei (like the posterior group, Po, or intralaminar/midline nuclei) originate from layers V and VI. In contrast to the layer V CT neurons, the layer VI CT neurons do innervate the TRN. The intrathalamic innervation pattern of layer VI CT neurons is regional whereas that of the layer V CT neurons, these latter CT neurons targeting only high-order thalamic nuclei, is more punctual. The layer V CT axon, which does not innervate the TRN, is a branch of the corticofugal main axon that targets the lower motor centers (brainstem and spinal cord). **(A2)** Design timeline illustrating the principal steps of the experiment. The color code of the brain state is dark gray for anesthesia, light gray for sedation and dark for death. **(B)** The left microphotograph shows, at low-magnification, the track left by the electrode and the extracellular labeling of the neurons located at and close to the recording site (here in the VL); the middle microphotograph shows, at higher-magnification, the recording site in the thalamic medial dorsal nucleus (MD, indicated by the arrowhead) and the somatodendritic complex of a couple of MD neurons; the left microphotograph shows the recording site with a few neural elements labeled in the TRN . On the right is presented, into a coronal plane, a mapping of the recording sites (black dots) into the TRN and the dorsal thalamic nuclei. The coronal plane represents a block of brain of about 3.6 mm thick posterior to the bregma (from -2.3 to -3.6 mm) in which recordings were performed. **(C)** Under the saline (control) condition, the cortex, the TRN and the dorsal thalamic nuclei exhibit a synchronized state, characterized by the occurrence of low-frequency (1-16 Hz) oscillations, including spindles. The extracellular TRN recordings can contain high-brain-frequency (200-700 Hz) bursts of APs (hfBurst, indicated by arrows). The framed expanded trace shows a couple of hfBursts associated with TC spindle waves. Under the ketamine condition, the TC system displays a more desynchronized state, characterized by the prominent occurrence of fast activities (>16 Hz), which include gamma-frequency oscillations. And the TRN cell fires more in the single AP (sAP) mode than in the hfBurst mode. Extracellular sAPs are indicated by the dots. Below, the expanded trace reveals sAPs associated with TC gamma waves. Single APs are also identifiable (indicated by dots) in the extracellular recording of the MD under the ketamine condition. **(D)** Spectral analysis of the cortical EEG (top) and of the thalamic extracellular activities (bottom) recorded under the saline then the ketamine conditions. Each value is a grand average (±SEM) from 6 rats, each rat being its control (per value: 23 epochs of 2 s/rat (hamming, resolution: 0.5 Hz)). In each chart, the part delimited by 2 dotted lines indicates the sigma-frequency band, which corresponds predominantly to spindles. **(E)** Time course of the power of, from top to bottom, gamma oscillations, spindles, and delta oscillations recorded simultaneously in the frontal cortex (FrCx), the TRN and in the medial dorsal (MD) nucleus before and after subcutaneous administrations of saline and ketamine (at 20 and 40 min, respectively).

## RESULTS

### Ketamine reduces thalamocortical spindles and delta-frequency oscillations

The recordings started about 2 h after the onset of the infusion of the pentobarbital containing regimen (Fig. 1A2), that is when the on-line spectral analysis revealed a stationary amount of spindles and slower oscillations (Fig. S2, Fig. 1C), which were qualitatively similar to those recorded during the natural non-REM sleep (Fig. S1B2). Multisite extracellular recordings were performed in the TRN, in midline, posterior, and ventral thalamic nuclei. From ~5 min after a subcutaneous administration of ketamine (2.5 mg/kg), the pattern of the cortical and thalamic baseline sleep activities was dramatically reduced in amplitude, supplanted by a more desynchronized pattern (Fig. 1C). Indeed, ketamine significantly decreased the spindle density (Fig. S3A), and the power (synchronization index) of the spindles and delta oscillations (Figs. 1D,E, S3B). It also decreased the amount of theta-frequency (5-9 Hz) oscillations (Fig. 1D), a CT theta activity that is a hallmark of drowsiness (Pinault et al., 2001). Concomitantly, ketamine significantly increased the power of ongoing gamma- and higher-frequency oscillations. The ketamine effects, observed in all recorded regions (n≥4 rats/region; Fig. 2), were transient (peaking at 15-20 min) with partial recovery at 60-80 min after the administration (Fig. 1E). In contrast to drugs modulating dopaminergic and cholinergic transmitter systems, dizocilpine (MK-801), a more specific NMDA receptor antagonist, well mimicked the ketamine effects on spindles and higher-frequency oscillations (S4). And the cholinomimetic physostigmine simulated the ketamine effects on delta oscillations and spindles, not on gamma and higher-frequency oscillations (Fig. S4a,b).

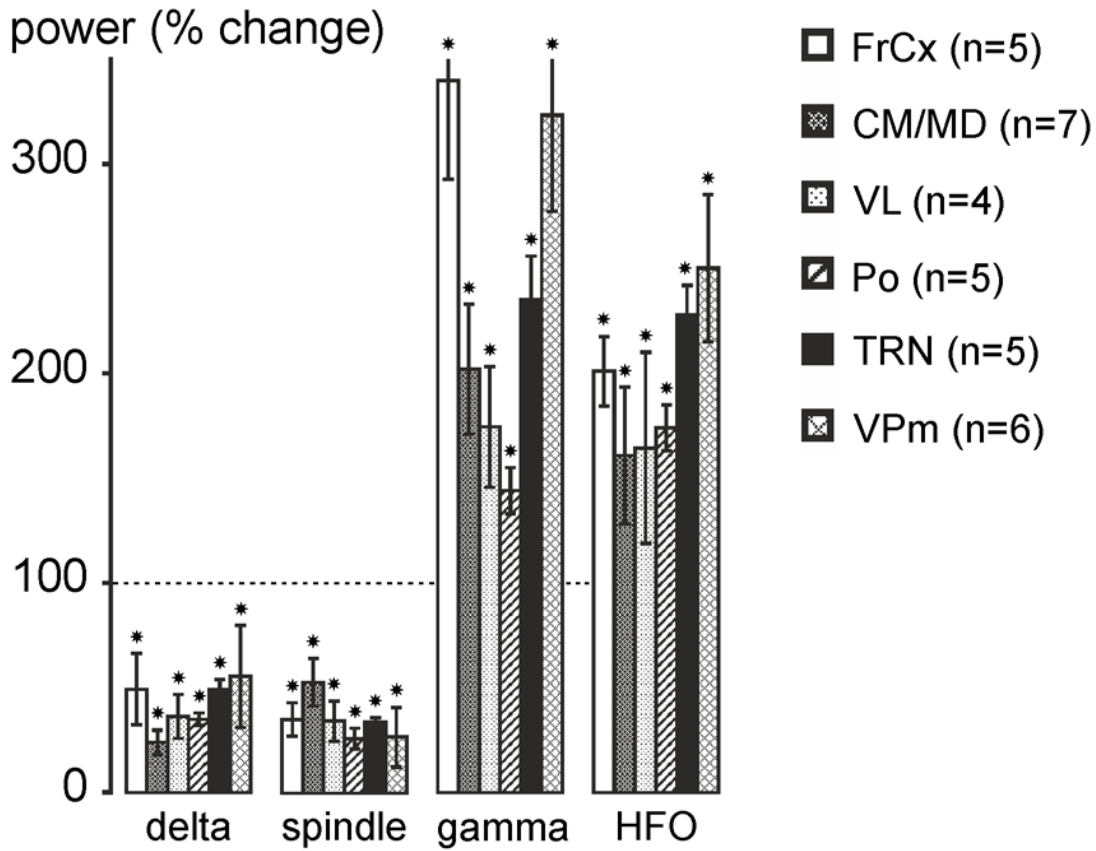

**Figure 2: Ketamine reduces delta-frequency oscillations and spindles and increases gamma- and higher-frequency oscillations.** The histogram shows the ketamine-induced percent changes (±SEM, relative to the saline condition, each rat being its control; post-ketamine: 20 to 30 min) in power of delta oscillations, spindles, gamma- and higher-frequency oscillations recorded in the frontal cortex, in the TRN and in first-order (VPm, VL) and in higher-order (CM/MD, Po) thalamic nuclei. Number of rats given in parentheses. Paired t-test relative to saline condition (star when p < 0.05). For abbreviations, see Fig. 1 legend.

## Ketamine switches the firing pattern of thalamic relay and reticular neurons from the burst mode to the tonic mode

In the following, all data are from the somatosensory system as it contains < 1% of local-circuit neurons (Harris and Hendrickson, 1987) and its 3-neuron layer VI CT-TRN-TC circuit, common to all nuclei of the dorsal thalamus, is the leading circuit in the generation of spindles. The location of the recording sites was identified based on electrophysiological and anatomical features (Fig. S5). From 11 extracellular thalamic recordings, 6 (from 6 rats) contained at least two TC units that were detectable using an automated spike sorting procedure (Fig. S6). Five out of 8 dual juxtacellular TC-TRN recordings (5 rats) had a duration long enough for data analyses under control and ketamine conditions, and 8 out of 15 TC cells met the intracellular requirements (Pinault et al., 2006).

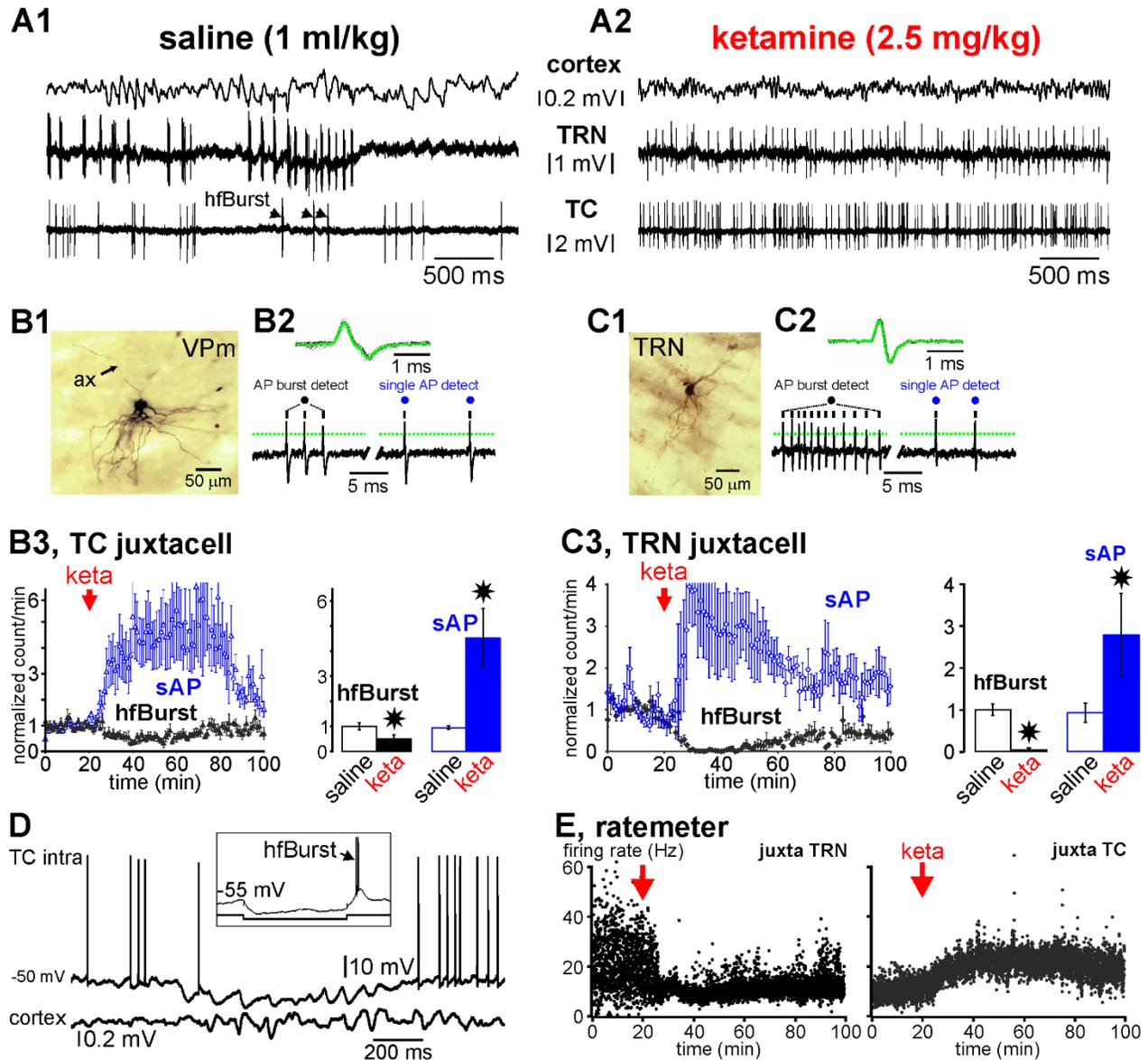

**Figure 3: Ketamine switches the firing pattern from a burst mode to a single action potential mode in thalamic relay (glutamatergic) and reticular (GABAergic) neurons.**

**(A1, A2)** Typical simultaneous recordings of the cortex (EEG), and of two single TRN and TC neurons (juxtacellular configuration) of the somatosensory system. Under the saline (A1, control) condition, the cortex displays a synchronized state, characterized by the occurrence of medium-voltage (>0.1 mV) low-frequency (1-16 Hz) oscillations, the TRN cell exhibits a typical series of rhythmic robust high-frequency bursts of action potentials (hfBursts, 300-500 APs/s), and the TC neuron exhibits single action potentials (sAPs) and, during the TRN burst series, a few bursts. A few minutes after the systemic administration of ketamine (A2, here: +20 min), the cortex displays a more desynchronized state, characterized by the prominent occurrence of lower voltage (<0.1 mV) and faster activities (>16 Hz), which include gamma-frequency oscillations. Under the ketamine condition, both the TC and the TRN cells exhibit much more sAPs than hfBursts. **(B1-B3)** Data from juxtacellularly recorded TC neurons. **(B1)** Photomicrography of parts of the somatodendritic complex and of the main axon (ax) of a juxtacellularly recorded and labeled (with Neurobiotin) TC neuron of the somatosensory thalamus. **(B2, top)** Average and superimposition of 50 action potentials. **(B2, below)**: Detection (from a voltage threshold, indicated by a dotted line) of a typical hfBurst of 3 APs and of 2 successive single APs. **(B3)** The density (number per minute, ±SEM, 5 TC cells from 5 rats) of hfBursts and of sAPs under the saline and ketamine conditions. Paired t-test (star when p<0.05). **(C1-C3)** Data from juxtacellularly recorded TRN neurons. **(C1)** Photomicrography of part of the somatodendritic complex of a juxtacellularly recorded and labeled (with Neurobiotin) TRN cell. **(C2, top)** Average and superimposition of 50 APs. **(C2, below)**: Detection (from a voltage threshold, indicated by a dotted line) of a typical hfBurst of 12 APs and of 2 successive single APs. **(C3)** The density (number per minute, ±SEM, 5 TRN cells from 5 rats) of hfBursts and of sAPs under the saline and ketamine conditions. Paired t-test (star when p<0.05). **(D)** Representative trace of an intracellularly recorded TC neuron showing the occurrence of subthreshold oscillations, including spindle-frequency rhythmic waves, which are concomitant with a synchronized EEG state in the related cortex. Note that the subthreshold oscillations occur during the through of a long-lasting hyperpolarization. In the frame is shown the occurrence of a low-threshold potential topped by a high-frequency burst of APs (hfBurst) at the offset of a 200-ms hyperpolarizing pulse. **(E)** Ratemeter of simultaneously juxtacellularly recorded TRN and TC neurons under saline then ketamine conditions. Each dot is the average (n = 5 neurons from 5 rats) of the number of inter-AP intervals per second.

**Thalamic relay neurons:** During the sedation, the extracellularly recorded TC units presented an irregular firing pattern consisting in hfBursts and single APs (Fig. S6A). It was extremely rare to see series of rhythmic hfBursts at the spindle frequency, suggesting that most of the TC spindle oscillations were subthreshold (Pinault et al., 2006), as demonstrated by the dual juxtacellular TRN-TC recordings (Fig. 3A1) and by the intracellular recordings of TC neurons (Fig. 3D). From ~5 min after the ketamine administration (16 TC units from 6 rats), the density of hfBursts significantly decreased whereas that of single APs increased for at least 60 min (Fig. S6B). The spike sorting method may, however, not be precise and reliable as the amplitude and shape of the APs might not be stationary over time (Lewicki, 1998). For instance, in TC hfBursts, the AP amplitude became progressively smaller (Fig. S6A).

Therefore, to better validate the ketamine effects observed in the extracellular TC recordings, we performed dual juxtacellular recordings of thalamic relay and reticular neurons. The juxtacellular single-unit recording-labeling technique allows the formal identification of the recorded neuron (Fig. 3A1,B1,C1) (Pinault, 1996). Ketamine, transiently and significantly, decreased the density of TC hfBursts and increased that of single APs (Fig. 3A1,A2 and B3). However, the decrease in the hfBurst density was ~50%, meaning that AP bursts still occurred under the ketamine condition. Embedded in the irregular tonic AP trains, a lot of them were doublets and triplets, whose intra-frequency was lower (inter-AP interval peak at 5-6 ms, Fig. 4A1,A2) than that of typical hfBursts (interval peak at 2-3 ms, Fig. 4A1). A partial recovery was noticeable 60-80 min after the ketamine administration (Fig. S6B, and Fig. 3B3). Of importance, ketamine increased the firing frequency band of TC neurons from, on average, 5-20 Hz (10.8±2.9 Hz, N=5 from 5 rats) to 15-30 Hz (21.7±3.5 Hz, N=5) (Fig. 3E). Furthermore, in one of the experiments, designed to record in the posterior group (equivalent to the pulvinar in humans) of the thalamus, 2 nearby (100 $\mu$m apart) TC cells were simultaneously recorded in the juxtacellular configuration (Fig. S7). Ketamine consistently augmented their firing frequency band in a similar way (from 0-10 Hz to 0-35 Hz).

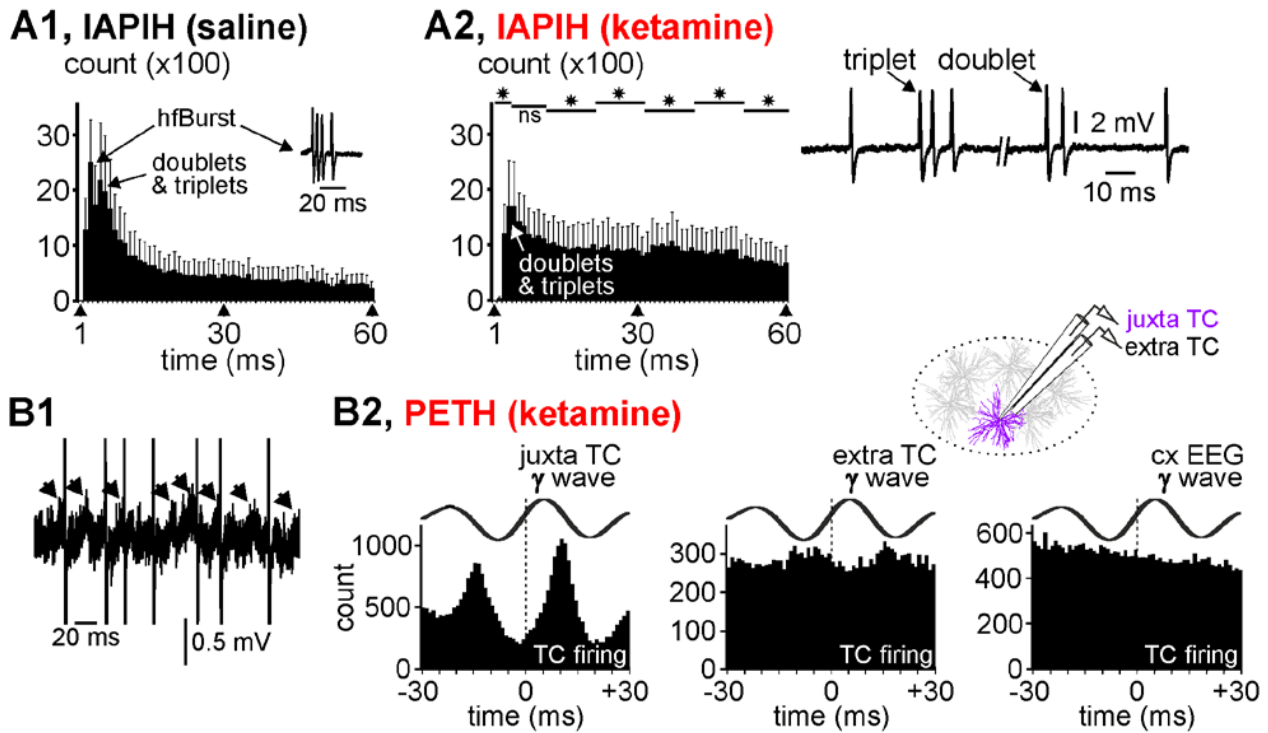

**Figure 4: Thalamocortical firing related to gamma-frequency oscillations. (A1, A2)** Averaged cumulated inter-AP interval histograms (IAPIH) from 5 juxtacellularly recorded TC cells (from 5 rats) under the control (A1) then the ketamine (A2) conditions (keta +15-25 min). Note the ketamine-induced diminution in the number of the short-lasting IAPIs, especially those composing high-frequency bursts of APs (IAPI = 2-10 ms). A typical TC hfBurst (IAPI at 2-3 ms) is shown in the control histogram. The remaining bursts were slower (IAPI at 5-6 ms) and shorter (e.g, especially doublets and triplets, like those shown on the right). Star when significant (paired t-test, p <0.05). **(B1)** A typical short-lasting trace of a juxtacellularly recorded TC cell showing low-amplitude gamma-frequency oscillations in the field potential and the AP occurrence at some cycles of the gamma oscillation. Each arrow indicates a juxtacellular gamma wave. The APs are truncated. **(B2)** Peri-event (gamma wave) time histogram (1-ms resolution) of the TC firing (cumulative count) under the ketamine condition (5 TC cells from 5 rats). Every gamma wave (juxta TC, extra TC (inter-tip distance = 100 μm, see drawing), and cxEEG) is an average of 100 filtered (25-55 Hz) individual gamma (γ) waves. Time "0" corresponds to the time at which gamma waves were detected.

Curiously, under the ketamine condition, the mean firing frequency of TC neurons (<30 Hz) was lower than the network gamma-frequency oscillations (frequency at maximal power: 33.6±1.1 Hz, n=7), raising the question whether or not TC single APs were related to the juxta- and extracellular gamma oscillations. In an attempt to address this question, firstly we looked at the raw juxtacellular recordings, in which we notice that TC neurons did not emit an AP at every wave of the gamma oscillations, which were not perfectly regular in waveform and timing (Figs. 4B1, S8), suggesting that the juxtacellular field potential variations reflected more membrane potential oscillations than APs. Secondly, a substantial number of single APs were phase-related to both the juxtacellular and the extracellular (100 μm apart) gamma waves (Fig. S8, Fig. 4B2). However, the temporal link was stronger with the juxtacellular (cellular activity) than the extracellular (nearby network activity) wave. In contrast to layer-organized cortical structures, the weak relation between the juxtacellular APs and the extracellular gamma waves seen in the somatosensory thalamus might have been due to an anarchic overlap of the current sinks and sources generated by the neural activities. On the other hand, there was no apparent relation between the TC firing and the cortical gamma waves (Fig. 4B2), which is not surprising as the EEG integrates the activities of interweaved large-scale networks.

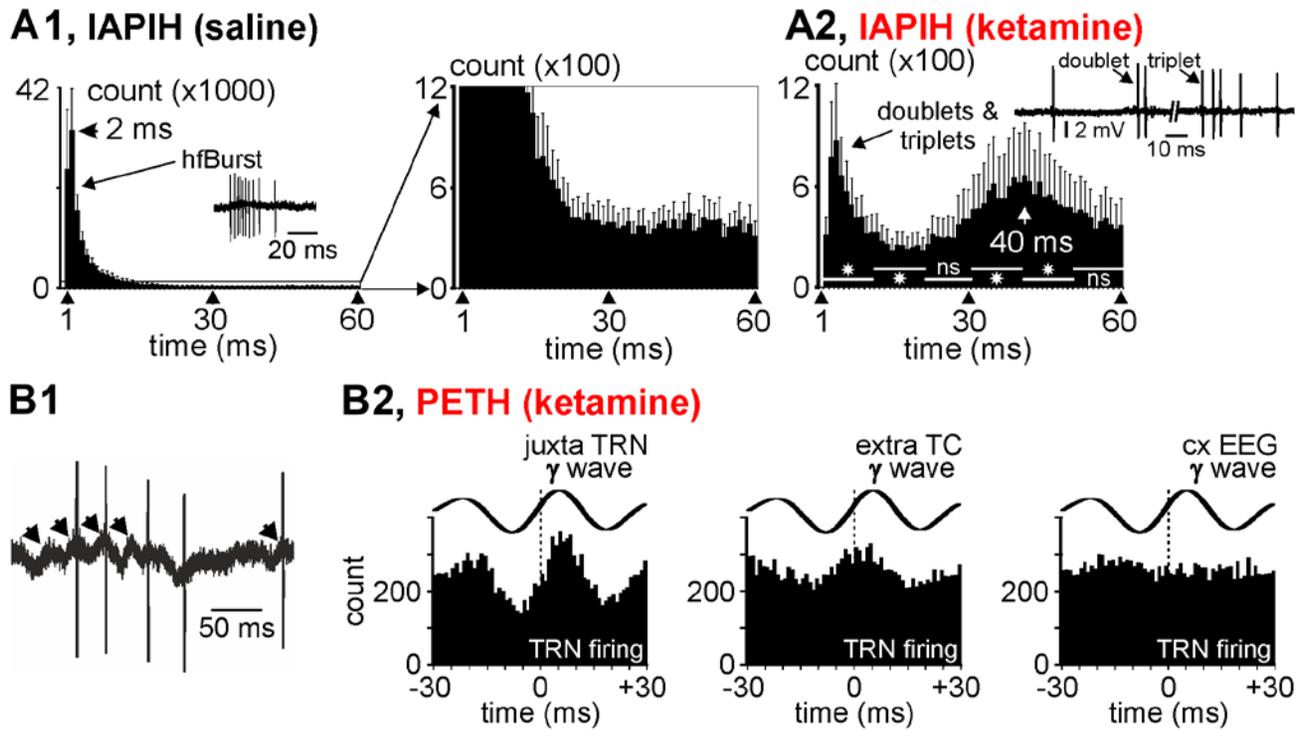

**Figure 5: Thalamic reticular nucleus firing related to gamma-frequency oscillations. (A1, A2)** Averaged cumulated inter-AP interval histograms (IAPIH) from 5 juxtacellularly recorded TRN cells (from 5 rats) under the control (A1) then the ketamine (A2) conditions (keta +15-–25 min). In (A1), are shown the control (saline) IAPIH in full (left) and partial (right) Y scales. Note the ketamine-induced dramatic diminution in the number of the short-lasting IAPIs, especially those composing high-frequency bursts of APs (IAPI = 2-–10 ms). A typical TRN hfBurst is shown in the control histogram. The remaining bursts were slower (increase in IAPIs) and shorter (e.g, especially doublets and triplets, like those shown). Star when significant (paired t-test, p < 0.05). **(B1)** A typical short-lasting trace of a juxtacellularly recorded TRN cell showing low-amplitude gamma-frequency oscillations in the field potential and that the AP occurrence at some cycles of the gamma oscillation. Each arrow indicates a juxtacellular gamma wave. **(B2)** Peri-event (gamma wave) time histogram of the TRN firing (cumulative count) under the ketamine condition (5 TRN cells from 5 rats). Every gamma wave (juxta TRN, extra TC, and cxEEG) is an average of 100 filtered (25-–55 Hz) individual gamma (γ) waves. Time "0" corresponds to the time at which gamma waves were detected.

**Thalamic reticular neurons:** During sedation, all extracellularly (Fig. 1C) or juxtacellularly (Fig. 3A1) recorded TRN cells exhibited sequences of rhythmic hfBursts in relation to the sleep TC oscillations. The burst sequence naturally recurred at a low frequency (<1 Hz) (Figs. S2B1 and S5B), during which rhythmic hfBursts occurred at the sigma (spindle)- and lower-frequency bands, including the delta band. The rhythmic character of spindle burst patterns was identifiable with an autocorrelation histogram (Fig. S2D). In TRN neurons, such sustained rhythmic burst activity involves the activation of NMDA receptors (Jacobsen et al., 2001). From ~5 min after a single ketamine administration, all juxtacellularly recorded TRN cells suddenly and transiently switched their ongoing rhythmic burst firing pattern to a sustained tonic, single AP firing pattern (Fig. 3A1,A2,C3). Furthermore, ketamine decreased their firing frequency band from 0-60 Hz (16.5±2.5 Hz) to 5-25 Hz (8.1±1.8 Hz; n=5 from 5 rats) (Fig. 3E). Remarkably and significantly, the single AP density increased whereas the hfBurst density decreased (Figs. 3C3, 5A2, B1-B2). In the inter-AP interval histogram, the first peak at 2-4 ms, a marker of hfBursts, disappeared almost completely. Under the ketamine condition, the first peak (3-6 ms) reflects longer inter-AP intervals which, like in TC neurons, are the signature of doublets and triplets embedded in the irregular tonic AP trains (Fig. 5B2).

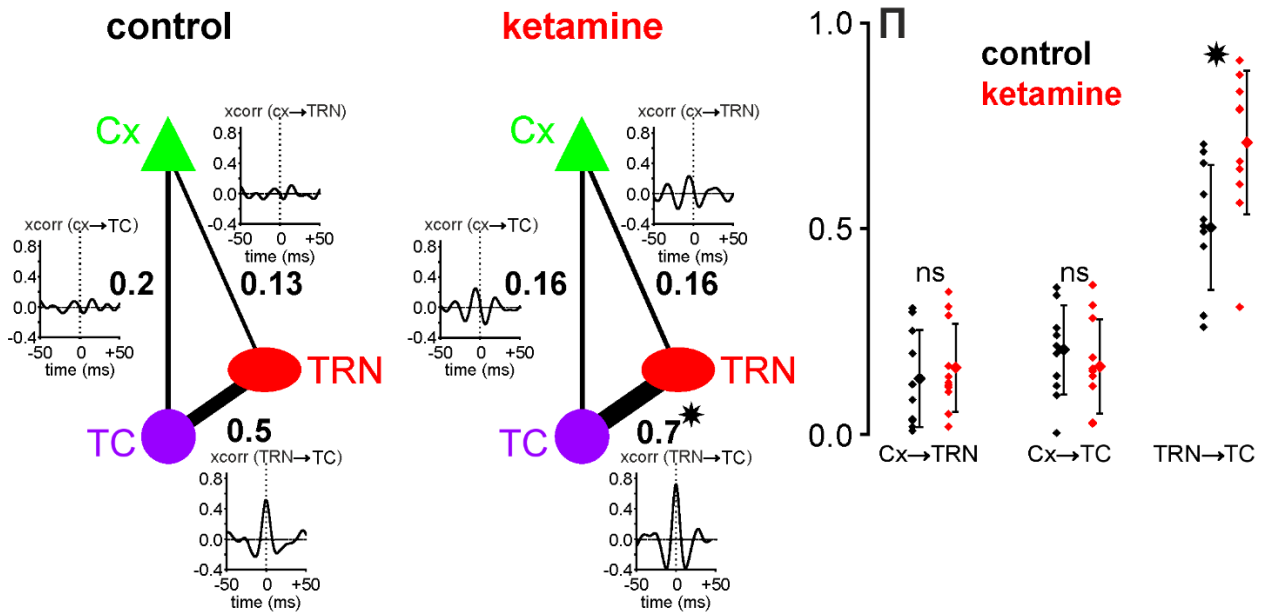

**Figure 6: Ketamine strengthens the functional gamma TRN-TC connectivity.** Direct interaction strength between two different sites is given by a partial correlation coefficient written (bold font) next to the edge connecting these sites. The plots presented next to these edges are cross-correlograms of 400ms-epochs from signals recorded at the corresponding sites and filtered in the gamma-range (25-55Hz). Under control condition, the strength of gamma interactions between TRN and TC sites was more than twice higher than for CT-TRN and CT-TC interactions, which was also reflected by a high peak in the TRN-TC cross-correlogram. When ketamine was applied, the strength of TRN-TC gamma interactions was significantly increased (paired t-test, p<0.001), resulting in a higher partial correlation coefficient and a higher peak in the average cross-correlogram. Although after ketamine application, correlations between CT and TRN and between CT and TC were higher in the cross-correlograms, the strength of CT-TRN and CT-TC interactions given by partial correlation coefficients did not change significantly (paired t-test, p>0.4). The plot on the right shows distributions of partial correlation coefficients *Π* for CT→TRN, CT→TC and TRN→TC gamma interactions in all experiments (N=11) under both control (black) and ketamine (red) conditions. (*) indicates significant difference revealed with a paired t-test with p<0.001; ns, non-significant.

Interestingly, the interval histogram reveals a second peak at ~30-50 ms. We predicted that the 30-50-ms peak represents a marker of juxtacellular gamma oscillations. Indeed, when looking closely at the juxtacellular recordings, it is obvious that the TRN cells fired at a certain proportion of gamma waves during their positive-going component (Fig. 5B1), meaning that the juxtacellular oscillations reflected threshold/suprathreshold and subthreshold membrane potential gamma oscillations. This observation is supported by a peri-gamma wave time histogram of the AP distribution (Fig5. B2), which shows that the probability of firing reached a maximum at (virtually 0 ms) the positive-going component of the gamma wave. Furthermore, a substantial number of TRN APs was also phase-related to gamma waves recorded extracellularly in the related somatosensory thalamic nucleus, suggesting a certain degree of functional connectivity. Moreover, using the partial correlation coefficient (S9), the strength of the gamma-frequency band TRN-TC connectivity was significantly increased by ketamine (Fig. 6). On the other hand, in the same way as TC neurons, there was no apparent relation between the TRN firing and the cortical gamma waves (Fig. 5B2).

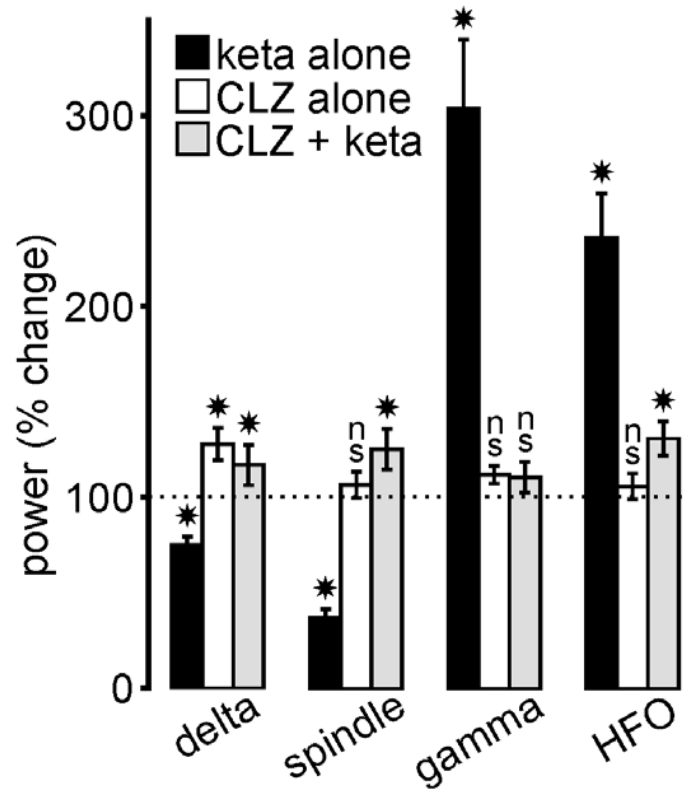

**Fig 7: Clozapine (CLZ) prevents the ketamine effects.** The histogram illustrates the drug-induced percent changes (mean±SEM; relative to their respective vehicle (saline for ketamine, saline/HCl 0.1N for clozapine) condition (100%, indicated by dotted line); 5 rats/condition) in power of all frequency bands in the cortical EEG. Student paired t-test: (*) p< 0.05; ns, not significant. In the CLZ + keta condition, ketamine (2.5 mg/kg) was administered 20 min after the CLZ (5 mg/kg) administration.

### Clozapine prevents the ketamine effects

Clozapine is one of the most effective antipsychotic drugs against treatment-resistant schizophrenia (Kane et al., 1988). Its clinical effects are thought to be related to interactions with a variety of receptors, including the glutamatergic receptors and more specifically NMDA receptors via the glycine site (Hunt et al., 2015; Lipina et al., 2005; Schwieler et al., 2008). Also, clozapine is well-known to modulate sleep spindles (Tsekou et al., 2015), possibly due to the activation of GABAergic TRN neurons via a specific action on D4 dopamine receptors (Mrzljak et al., 1996), which would exert a tonic influence on the TRN activity (Barrientos et al., 2019). Therefore, it was interesting to probe whether a single systemic administration of clozapine could prevent the ketamine effects on TC oscillations. To address this issue, clozapine was subcutaneously administered at a dose (5 mg/kg) that durably decreases the power of spontaneously-occurring cortical gamma oscillations in the naturally-behaving rat (Jones et al., 2012) 20 or 120 min before the ketamine challenges. In all rats (n=7), clozapine consistently prevented the ketamine peak (at ~15-20 min) effect on spindles, delta- and gamma-/higher-frequency oscillations (Fig. S10 and Fig. 7). When administered alone, clozapine significantly increased the power of delta oscillations (Fig. 7).

## DISCUSSION

In the present study conducted in the sedated rat, the psychotomimetic ketamine induced a transient dramatic decrease in ongoing thalamic and cortical spindles and slower oscillations, and a concomitant increase in gamma-/higher-frequency oscillations, which is reminiscent of an arousal effect. These new preclinical findings support the hypothesis of a reduced function of NMDA receptors in the reduction of spindles and slow-waves in schizophrenia. They also support the hypothesis that the spindle reduction observed in patients with schizophrenia is due to deficient TRN inhibition (Baran et al., 2019; Ferrarelli and Tononi, 2011; Manoach and Stickgold, 2019; Pratt and Morris, 2015; Young and Wimmer, 2017). However, because ketamine impacted simultaneously and similarly the cellular and network activity patterns of both the GABAergic TRN and the glutamatergic TC neurons (present study), and because ketamine disrupts the function of the CT pathway (Anderson et al., 2017) and forces the brain to generate persistent and generalized aberrant gamma oscillations in cortical and subcortical structures (Hakami et al., 2009; Slovik et al., 2017), non-exclusive alternative theories will be discussed in the following in an attempt to understand the possible underlying mechanisms.

### The arousal promoting effect of low-dose ketamine

Interestingly, the present findings are in agreement with a previous study performed in the awake rat demonstrating that, in a subpopulation of cortical neurons, NMDA receptor hypofunction produces a sustained increase in the firing rate (sAP mode) and a concomitant reduction of burst activity associated with a psychosis-relevant behavior (Jackson et al., 2004).

The arousal promoting effect of a single psychotomimetic dose of ketamine has been well documented in awake, free-behaving rats (Ahnaou et al., 2017; Hakami et al., 2009; Pinault, 2008). The ketamine-induced transient arousal effect is characterized by an abnormal, erratic behavior with hyperlocomotion, ataxias and stereotypies associated with deficits in cognitive performances (Chrobak et al., 2008; Pitsikas et al., 2008), and an excessive amplification of gamma-frequency oscillations (Ehrlichman et al., 2009; Hakami et al., 2009; Pinault, 2008). These transient, behavioral, cognitive and electrophysiological abnormalities are reminiscent of a psychosis-relevant behavior, during which not a single sleep episode was observed during the time dedicated to sleep (S1). Moreover, a comprehensive study demonstrated that ketamine delays the sleep onset latency (Ahnaou et al., 2017).

Clinical investigation showed that patients with psychosis have difficulties initiating sleep (Poulin et al., 2003). Abnormal levels of arousal may be a predictor of psychotic disorders (Lee et al., 2012; Tieges et al., 2013). Here, it is further shown that, in the sedated rat, ketamine elicited a fleeting arousal-like reaction, at least in the TC-TRN system, which is electrophysiologically reminiscent of REM sleep, a brain state considered as a natural model of psychosis (Dresler et al., 2015; Hobson, 1997; Mason and Wakerley, 2012; Mota et al., 2016; Scarone et al., 2008). Moreover, the NMDA receptor hypofunction-related increase in gamma-/higher-frequency oscillations observed in sedated rats is also recorded during the natural REM sleep (Kocsis, 2012b). So, we interpret the ketamine-induced desynchronized state as uncharacteristic REM-like sleep phenomena or a pathological persistent UP state (Fig. 8). During the ketamine-induced pathological UP state, expected to occur within diverse cortical and subcortical structures (Hakami et al., 2009), cortical and thalamic neurons would be more depolarized than during the DOWN state to generate more threshold (for AP initiation) and supra-threshold membrane potential oscillations (Destexhe and Pare, 1999). In thalamic neurons, the burst mode is a reliable hallmark of sleep oscillations, every hfBurst occurring at the top of a low-threshold $Ca^{++}$ potential mediated by the activation of T-type channels, which are de-inactivated via membrane hyperpolarization (<-60 mV) (Crunelli et al., 2006). Both the synaptic interactions between TC and TRN neurons and the intrinsic pacemaker properties of TRN cells are well-known to play leading roles in the generation of thalamic spindles (Steriade et al., 1985; Steriade et al., 1993). Under the ketamine condition, the substantial increase in the single AP density suggests that the membrane potential of TC and TRN neurons was more often depolarized. This is supported first by the occurrence of gamma oscillations and single AP firing in our juxtacellular TC and TRN recordings and, second, by an increase in the gamma band TRN-TC connectivity. The single AP mode is usually recorded when T-type $Ca^{++}$ channels are inactivated via membrane depolarization (>-60 mV) (Mulle et al., 1986). Disruption of the $Ca_V3.3$ $Ca^{++}$ channel, which encodes the low-threshold T channels (Astori et al., 2011), may be involved in the etio-pathophysiology of schizophrenia (Andrade et al., 2016).

**Contribution of the corticothalamic pathway.**

In the thalamus, ketamine would act principally on both the glutamatergic TC and the GABAergic TRN neurons. How did ketamine convert the firing from burst to the tonic mode simultaneously in both TC and TRN neurons? During sleep, sustained hyperpolarization would be the result of either excess inhibition or disfacilitation. Under the ketamine condition, a likely effect would be a sustained excitation of these two types of neurons by common afferent input. In addition to the influence of neuromodulatory inputs (see below), the CT pathway seems an excellent candidate (Lam and Sherman, 2010; Landisman and Connors, 2007). Indeed, the primary axon of the CT neurons splits into two branches, one innervating TRN neurons, the other TC neurons (Bourassa et al., 1995; Golshani et al., 2001). Furthermore, it is known that cortical GABAergic interneurons are highly sensitive to NMDA receptor antagonists (Grunze et al., 1996). The ketamine-induced NMDA receptor-mediated disfacilitation of the GABAergic cortical interneurons would be responsible for the disinhibition (or excitation) of glutamatergic pyramidal neurons (Homayoun and Moghaddam, 2007), including CT neurons. So, the disinhibition of CT neurons would lead to the generation of a sustained thalamic AMPA-mediated gamma hyperactivity (Anderson et al., 2017; Crandall et al., 2015; Golshani et al., 2001). And the ketamine-induced hyperactivation of layer VI CT neurons could in addition promote the gamma-frequency pacemaker properties of the GABAergic TRN cells (Pinault and Deschênes, 1992a, b). NMDA receptors are more critical for the CT-mediated excitation of TRN than TC neurons (Deleuze and Huguenard, 2016). Furthermore, the long-lasting kinetics of NMDA receptors in the GABAergic TRN neurons are essential to promote rhythmic $Ca^{++}$-mediated burst firing, which then cyclically hyperpolarizes the postsynaptic TC neurons through the activation of GABA receptors. Importantly, in TRN neurons, the NMDA-mediated effects of CT transmission can work across a wide range of voltages so as the voltage-dependent blockade by $Mg^{++}$ is incomplete and that NMDA receptors can be activated by synaptically released glutamate even in the absence of AMPA receptor-mediated activation (Deleuze and Huguenard, 2016). Thus, because CT neurons outnumber by a factor of ~10 TC neurons (Sherman and Koch, 1986), the ketamine NMDA-mediated effects are expected to be stronger on TRN than on TC neurons in reducing burst activity, which, in fact, was indeed observed in the present study (Fig3B3,C3). The NMDA receptor hypofunction-related spindle reduction and gamma increase in the TRN-TC system may help to understand the increased TC connectivity correlated with spindle deficits in schizophrenia (Baran et al., 2019).

Ketamine, at a psychotomimetic dose, is expected to affect almost, if not all, brain neurons. In the thalamus, it would impact at least TC and TRN neurons, which work together because of reciprocal connections through open and closed-loop circuits (Pinault and Deschenes, 1998). Interestingly, under the ketamine condition, both TRN and TC neurons fired in the single AP mode, and the TRN cells on average 2.7 times less than TC neurons. This finding is in line with an intra-cortical study showing that NMDA receptor hypofunction decreases the firing of GABAergic neurons and increases that of glutamatergic neurons (Homayoun and Moghaddam, 2007). Thus, NMDA receptor hypofunction would lead to TC and cortical excitations by disinhibition of the glutamatergic neurons, which would lead to an excessive accumulation of synaptic glutamate and subsequently to activation of AMPA receptors (Moghaddam et al., 1997). Such a psychosis-relevant state may be the source for the generation of abnormal internally-generated information (Gandal et al., 2012; Hakami et al., 2009).

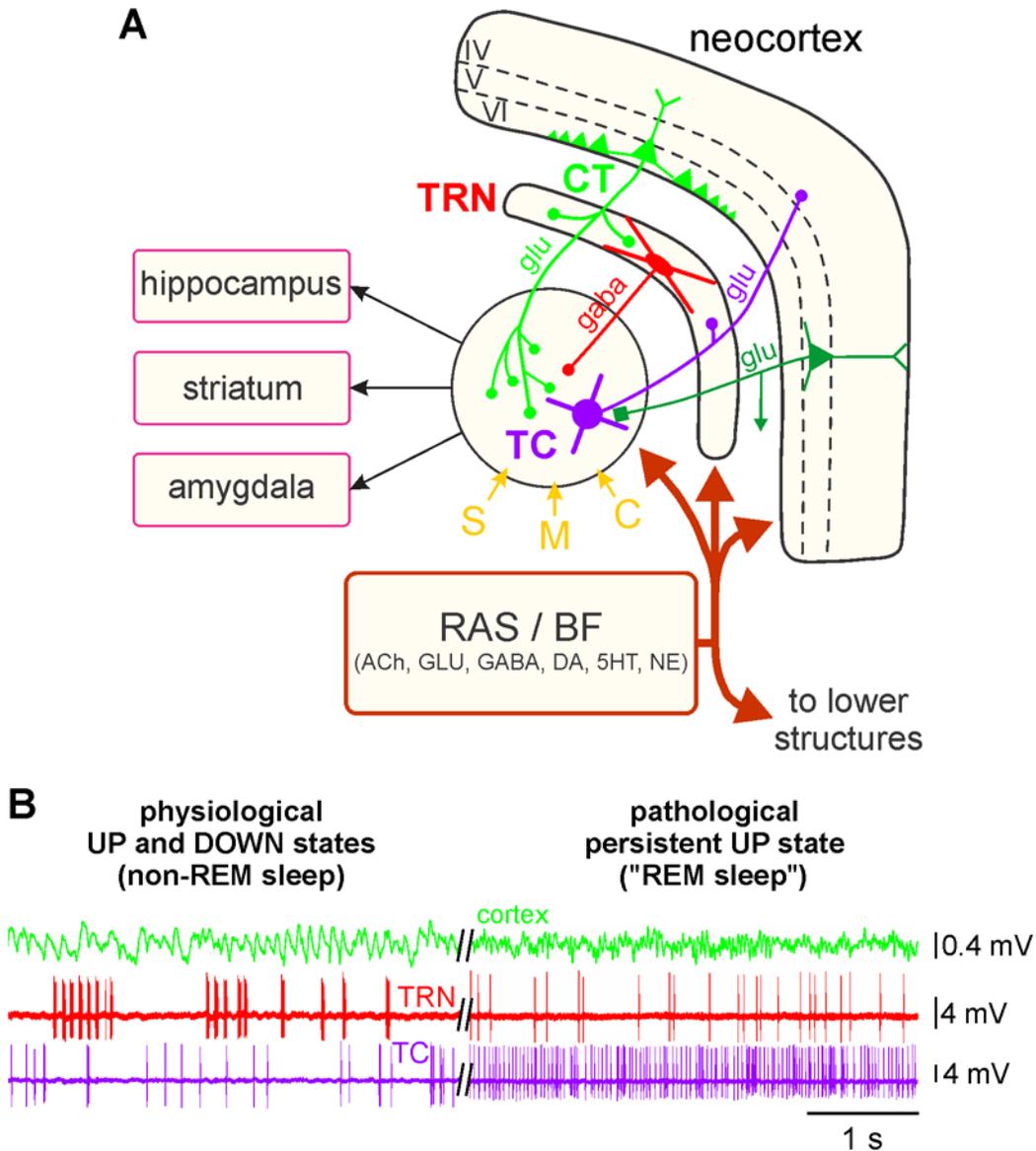

**Figure 8. Theoretical prediction of the ketamine action in both the ascending reticular activating system and the corticothalamic pathway. (A)** Simplified drawing of the hodology of the 4-neuron CT-TRN-TC circuit, which is considered as being the upper part of the ascending reticular activating system (RAS). See main text and Fig. 1 legend for detailed description of the circuit, which receives sensory (S), motor (M) and cognitive/associative (C) inputs. It is important to specify that the layer VI CT neurons outnumber by a factor of about 10 the TC neurons. The first- and higher-order thalamic nuclei are under the neuromodulatory influence of the various inputs from the ascending RSA and the basal forebrain (BF). **(B, left)** Physiological UP and DOWN states: During the non-REM sleep, the TC system displays principally a synchronized state, characterized by the occurrence of delta oscillations and spindles; the TRN cell exhibits mainly rhythmic (at the delta-, theta- and spindle-frequency bands) hfBursts of action potentials. The synchronized state includes two sub-states, UP and DOWN, which are usually associated with active and quiescent cellular firings, respectively. **(B, right)** Pathological persistent UP state: This ketamine-induced persistent UP state is assumed to be an abnormal REM sleep. After a single systemic administration of a subanesthetizing low-dose of ketamine, the TC system displays a more desynchronized state (peak effect at about +15-20 min) characterized by the prominent occurrence of lower voltage and faster activities (>16 Hz), which include beta-, gamma- and higher-frequency oscillations. Under the ketamine condition, both the TC and the TRN neurons exhibit a persistent irregular and tonic firing containing more single APs than hfBursts. ACh, acetylcholine; GLU, glutamate; 5HT, serotonin; DA, dopamine; NE, norepinephrine.

**Contribution of the ascending reticular activating system and basal forebrain**

The likely mechanisms underlying the effects of ketamine remain debatable as it acts in all brain structures and at many receptors (Dorandeu, 2013; Sleigh et al., 2014; Zanos et al., 2018). The ketamine-induced acute arousal-like effect may involve, among many others, cholinergic, monoaminergic, and orexinergic arousal systems (Ahnaou et al., 2017; Dawson et al., 2013; Lu et al., 2008). We should not exclude that, under our experimental conditions, the observed physostigmine effects (decrease in spindles and slower waves) suggest that ketamine could have acted also at acetylcholine receptors. Interestingly, the fact that a single low-dose of ketamine simultaneously affected, in an opposite manner, spindle-/delta-frequency and gamma-/higher-frequency TC oscillations is reminiscent of the seminal finding of Moruzzi and Magoun (Moruzzi and Magoun, 1949). Indeed, these pioneering investigators demonstrated that electrical stimulation of the reticular formation, a complex set of interconnected circuits within the brainstem, evokes in the TC system a switch of the EEG pattern from a synchronized to a desynchronized state, an effect interpreted as an EEG arousal reaction. Moreover, activation of the mesencephalic reticular formation effectively desynchronizes the cortical EEG in lighly anesthetized animals (Munk et al., 1996). Thus, the present findings give further support to the hypothesis of a dysregulation of the ascending reticular activating system, which includes the pedunculopontine nucleus, the basal forebrain, and the thalamus, in the etio-pathophysiology of psychotic disorders (Dawson et al., 2013; Garcia-Rill et al., 2015; Heimer, 2000; Howland, 1997). Moreover, in a previous investigation we demonstrated that NMDA receptor hypofunction leads to a persistent increase in gamma oscillations in the basalis, a cholinergic structure with widespread axonal projections well-known to modulate the neocortex and the TC-TRN system (Hakami et al., 2009; Pinault and Deschenes, 1992).

Also, for the observed ketamine effects, we should not exclude a contribution of the ascending GABAergic pathways (originating from the brainstem, midbrain, ventral tegmental area, zona incerta, basal ganglia, and from the basal forebrain), which play a critical role in promoting TC activation, arousal and REM sleep (Brown and McKenna, 2015; Kim et al., 2015). In the present study, interestingly, physostigmine, known to promote REM sleep (Sitaram et al., 1976) and a cortical EEG arousal (Kenny et al., 2016; Roy and Stullken, 1981), exerted a ketamine-like effect on delta oscillations and spindles (Fig. S4a,b) and on the firing pattern of TRN neurons (Fig. S4c). The atypical antipsychotic clozapine consistently prevented the foremost ketamine-induced acute effects on sleep oscillations. The fact that clozapine alone increased delta oscillations, but did not enhance sleep spindles or reduce gamma activity may indicate a general slowing of the EEG power, which may counteract the arousal effects of ketamine and help keeping the rats in the slow wave sleep (Hinze-Selch et al., 1997). The fact that, in contrast to ketamine, clozapine alone did not affect the cortical gamma power suggests that ketamine and clozapine exerted their action via distinct neural/molecular targets, which does not discredit the hypothesis of a dysregulation of the reticular activating system (Dawson et al., 2013; Garcia-Rill et al., 2015; Heimer, 2000; Howland, 1997), the TC system being nothing but its downstream part (Fig. 8).

Further investigation is required to better understand, in the thalamus, the mechanisms underlying the relative contribution of the top-down and bottom-up effects of low-dose ketamine.

### Conclusions and significance

The present preclinical investigation with its limitations (S11) demonstrates that the acute effects of ketamine result in fast onset arousal promoting effect, suggesting that it acts like a rapid-acting inducer of REM sleep-associated cognitive processes, which is reminiscent of its ability to induce hallucinatory and delusional symptoms (Baldeweg et al., 1998; Becker et al., 2009; Behrendt, 2003; Ffytche, 2008; Spencer et al., 2004). Low-dose ketamine not only disturbs brain rhythms, but also disrupts attention-related sensorimotor and cognitive processes (Grent-'t-Jong et al., 2018; Hoflich et al., 2015; Hong et al., 2010), supporting the notion that schizophrenia is a cognitive disorder with psychosis as a subsequent consequence (Cohen and Insel, 2008; Huang et al., 2019; Woodward and Heckers, 2016). The ketamine-induced changes in rodent EEG oscillations are reminiscent of those observed in at-risk mental state individuals (Fleming et al., 2019; Ramyead et al., 2015) and during the first episode of schizophrenia (Andreou et al., 2015; Flynn et al., 2008). Taken together, the present findings support more strongly the whole brain-networks hypothesis than the isolated brain circuit theory of schizophrenia (Kambeitz et al., 2016).

The neural mechanisms underlying the ketamine-induced fleeting arousal-like effect may be, in part, those responsible for the initial stage of the rapid-acting antidepressant action of ketamine in patients with drug-resistant major depressive disorders (Duncan et al., 2019; Krystal et al., 2019; Nugent et al., 2019), leading us to think that the ketamine effects are state-dependent. In addition, the present results suggest that the combined sleep and ketamine models have some predictive validity for the first-stage development of innovative therapies against psychotic, bipolar, and depressive disorders.

## ACKNOWLEDGEMENTS

The present work was supported by INSERM, the French National Institute of Health and Medical Research (Institut National de la Santé et de la Recherche Médicale, 2013-), l'Université de Strasbourg, Unistra (2013-), and Neurex. This project has been funded with support from the NeuroTime Erasmus+ program of the European Commission (2015-2020: AM and YQ). This publication reflects the views only of the authors, and the Commission cannot be held responsible for any use which may be made of the information contained therein. ASA is a graduate student from the Euridol Graduate School of Pain. Data of the present study were presented in 2018 at both the FENS Forum (Berlin) and the SFN meeting (San Diego). The authors thank Yoland Smith and Martin Deschênes for critical reading of the manuscript.

## DISCLOSURES

All authors have approved the final version of the article. The authors report no competing biomedical financial interests or potential conflicts of interest.

## AUTHORS' CONTRIBUTION

AM, YQ, SK, DP: Design, data acquisition & analysis, and writing; ASA: Data acquisition & analysis; DC: Animal well-being, surgery and technical aspects.

## REFERENCES

Ahnaou, A., Huysmans, H., Biermans, R., Manyakov, N.V., Drinkenburg, W., 2017. Ketamine: differential neurophysiological dynamics in functional networks in the rat brain. Transl Psychiatry 7(9), e1237.

Anderson, P.M., Jones, N.C., O'Brien, T.J., Pinault, D., 2017. The N-Methyl d-Aspartate Glutamate Receptor Antagonist Ketamine Disrupts the Functional State of the Corticothalamic Pathway. Cereb Cortex 27(6), 3172-3185.

Andrade, A., Hope, J., Allen, A., Yorgan, V., Lipscombe, D., Pan, J.Q., 2016. A rare schizophrenia risk variant of CACNA1I disrupts CaV3.3 channel activity. Sci Rep 6, 34233.

Andreasen, N.C., 1997. The role of the thalamus in schizophrenia. Can J Psychiatry 42(1), 27-33.

Andreou, C., Nolte, G., Leicht, G., Polomac, N., Hanganu-Opatz, I.L., Lambert, M., Engel, A.K., Mulert, C., 2015. Increased Resting-State Gamma-Band Connectivity in First-Episode Schizophrenia. Schizophr Bull 41(4), 930-939.

Anticevic, A., Corlett, P.R., Cole, M.W., Savic, A., Gancsos, M., Tang, Y., Repovs, G., Murray, J.D., Driesen, N.R., Morgan, P.T., Xu, K., Wang, F., Krystal, J.H., 2015. N-methyl-D-aspartate receptor antagonist effects on prefrontal cortical connectivity better model early than chronic schizophrenia. Biol. Psychiatry 77(6), 569-580.

Astori, S., Wimmer, R.D., Prosser, H.M., Corti, C., Corsi, M., Liaudet, N., Volterra, A., Franken, P., Adelman, J.P., Luthi, A., 2011. The Ca(V)3.3 calcium channel is the major sleep spindle pacemaker in thalamus. Proc Natl Acad Sci U S A 108(33), 13823-13828.

Bal, T., Debay, D., Destexhe, A., 2000. Cortical feedback controls the frequency and synchrony of oscillations in the visual thalamus [In Process Citation]. J Neurosci 20(19), 7478-7488.

Baldeweg, T., Spence, S., Hirsch, S.R., Gruzelier, J., 1998. Gamma-band electroencephalographic oscillations in a patient with somatic hallucinations. Lancet 352(9128), 620-621.

Baran, B., Karahanoglu, F.I., Mylonas, D., Demanuele, C., Vangel, M., Stickgold, R., Anticevic, A., Manoach, D.S., 2019. Increased Thalamocortical Connectivity in Schizophrenia Correlates With Sleep Spindle Deficits: Evidence for a Common Pathophysiology. Biol Psychiatry Cogn Neurosci Neuroimaging 4(8), 706-714.

Barrientos, R., Alatorre, A., Martinez-Escudero, J., Garcia-Ramirez, M., Oviedo-Chavez, A., Delgado, A., Querejeta, E., 2019. Effects of local activation and blockade of dopamine D4 receptors in the spiking activity of the reticular thalamic nucleus in normal and in ipsilateral dopamine-depleted rats. Brain Res 1712, 34-46.

Becker, C., Gramann, K., Muller, H.J., Elliott, M.A., 2009. Electrophysiological correlates of flicker-induced color hallucinations. Conscious. Cogn 18(1), 266-276.

Behrendt, R.P., 2003. Hallucinations: synchronisation of thalamocortical gamma oscillations underconstrained by sensory input. Conscious. Cogn 12(3), 413-451.

Bonjean, M., Baker, T., Lemieux, M., Timofeev, I., Sejnowski, T., Bazhenov, M., 2011. Corticothalamic feedback controls sleep spindle duration in vivo. J Neurosci 31(25), 9124-9134.

Bourassa, J., Pinault, D., Deschênes, M., 1995. Corticothalamic projections from the cortical barrel field to the somatosensory thalamus in rats: a single-fibre study using biocytin as an anterograde tracer. Eur J Neurosci 7(1), 19-30.

Brown, R.E., McKenna, J.T., 2015. Turning a Negative into a Positive: Ascending GABAergic Control of Cortical Activation and Arousal. Front Neurol 6, 135.

Castelnovo, A., Graziano, B., Ferrarelli, F., D'Agostino, A., 2017. Sleep spindles and slow waves in schizophrenia and related disorders: main findings, challenges and future perspectives. Eur J Neurosci.

Chrobak, J.J., Hinman, J.R., Sabolek, H.R., 2008. Revealing past memories: proactive interference and ketamine-induced memory deficits. J. Neurosci 28(17), 4512-4520.

Clinton, S.M., Meador-Woodruff, J.H., 2004a. Abnormalities of the NMDA Receptor and Associated Intracellular Molecules in the Thalamus in Schizophrenia and Bipolar Disorder. Neuropsychopharmacology 29(7), 1353-1362.

Clinton, S.M., Meador-Woodruff, J.H., 2004b. Thalamic dysfunction in schizophrenia: neurochemical, neuropathological, and in vivo imaging abnormalities. Schizophr. Res 69(2-3), 237-253.

Cohen, J.D., Insel, T.R., 2008. Cognitive neuroscience and schizophrenia: translational research in need of a translator. Biol Psychiatry 64(1), 2-3.

Connor, L., Burrows, P.E., Zurakowski, D., Bucci, K., Gagnon, D.A., Mason, K.P., 2003. Effects of IV pentobarbital with and without fentanyl on end-tidal carbon dioxide levels during deep sedation of pediatric patients undergoing MRI. AJR Am J Roentgenol 181(6), 1691-1694.

Coyle, J.T., 2012. NMDA receptor and schizophrenia: a brief history. Schizophr Bull 38(5), 920-926.

Crandall, S.R., Cruikshank, S.J., Connors, B.W., 2015. A corticothalamic switch: controlling the thalamus with dynamic synapses. Neuron 86(3), 768-782.

Cronenwett, W.J., Csernansky, J., 2010. Thalamic pathology in schizophrenia. Curr. Top. Behav. Neurosci 4, 509-528.

Crunelli, V., Cope, D.W., Hughes, S.W., 2006. Thalamic T-type Ca2+ channels and NREM sleep. Cell Calcium 40(2), 175-190.

Dawson, N., Morris, B.J., Pratt, J.A., 2013. Subanaesthetic ketamine treatment alters prefrontal cortex connectivity with thalamus and ascending subcortical systems. Schizophr Bull 39(2), 366-377.

Deleuze, C., Huguenard, J.R., 2016. Two classes of excitatory synaptic responses in rat thalamic reticular neurons. J Neurophysiol 116(3), 995-1011.

Deschênes, M., Hu, B., 1990. Electrophysiology and pharmacology of the corticothalamic input to lateral thalamic nuclei: An intracellular study in the cat. Eur J Neurosci 2, 140-152.

Destexhe, A., Pare, D., 1999. Impact of network activity on the integrative properties of neocortical pyramidal neurons in vivo. J Neurophysiol 81(4), 1531-1547.

Dorandeu, F., 2013. Happy 50th anniversary ketamine. CNS Neurosci Ther 19(6), 369.

Dresler, M., Wehrle, R., Spoormaker, V.I., Steiger, A., Holsboer, F., Czisch, M., Hobson, J.A., 2015. Neural correlates of insight in dreaming and psychosis. Sleep Med Rev 20, 92-99.

Duncan, W.C., Jr., Ballard, E.D., Zarate, C.A., 2019. Ketamine-Induced Glutamatergic Mechanisms of Sleep and Wakefulness: Insights for Developing Novel Treatments for Disturbed Sleep and Mood. Handb Exp Pharmacol 253, 337-358.

Ehrlichman, R.S., Gandal, M.J., Maxwell, C.R., Lazarewicz, M.T., Finkel, L.H., Contreras, D., Turetsky, B.I., Siegel, S.J., 2009. N-methyl-d-aspartic acid receptor antagonist-induced frequency oscillations in mice recreate pattern of electrophysiological deficits in schizophrenia. Neuroscience 158(2), 705-712.

Ferrarelli, F., Huber, R., Peterson, M.J., Massimini, M., Murphy, M., Riedner, B.A., Watson, A., Bria, P., Tononi, G., 2007. Reduced sleep spindle activity in schizophrenia patients. Am J Psychiatry 164(3), 483-492.

Ferrarelli, F., Peterson, M.J., Sarasso, S., Riedner, B.A., Murphy, M.J., Benca, R.M., Bria, P., Kalin, N.H., Tononi, G., 2010. Thalamic dysfunction in schizophrenia suggested by whole-night deficits in slow and fast spindles. Am J Psychiatry 167(11), 1339-1348.

Ferrarelli, F., Tononi, G., 2011. The thalamic reticular nucleus and schizophrenia. Schizophr Bull 37(2), 306-315.

Ffytche, D.H., 2008. The hodology of hallucinations. Cortex 44(8), 1067-1083.

Fleming, L.M., Javitt, D.C., Carter, C.S., Kantrowitz, J.T., Girgis, R.R., Kegeles, L.S., Ragland, J.D., Maddock, R.J., Lesh, T.A., Tanase, C., Robinson, J., Potter, W.Z., Carlson, M., Wall, M.M., Choo, T.H., Grinband, J., Lieberman, J., Krystal, J.H., Corlett, P.R., 2019. A multicenter study of ketamine effects on functional connectivity: Large scale network relationships, hubs and symptom mechanisms. Neuroimage Clin 22, 101739.

Flynn, G., Alexander, D., Harris, A., Whitford, T., Wong, W., Galletly, C., Silverstein, S., Gordon, E., Williams, L.M., 2008. Increased absolute magnitude of gamma synchrony in first-episode psychosis. Schizophr Res 105(1-3), 262-271.

Gandal, M.J., Edgar, J.C., Klook, K., Siegel, S.J., 2012. Gamma synchrony: towards a translational biomarker for the treatment-resistant symptoms of schizophrenia. Neuropharmacology 62(3), 1504-1518.

Ganes, T., Andersen, P., 1975. Barbiturate spindle activity in functionally corresponding thalamic and cortical somato-sensory areas in the cat. Brain Res 98(3), 457-472.

Garcia-Rill, E., D'Onofrio, S., Mahaffey, S., Bisagno, V., Urbano, F.J., 2015. Pedunculopontine arousal system physiology-Implications for schizophrenia. Sleep Sci 8(2), 82-91.

Gentet, L.J., Ulrich, D., 2003. Strong, reliable and precise synaptic connections between thalamic relay cells and neurones of the nucleus reticularis in juvenile rats. J Physiol 546(Pt 3), 801-811.

Golshani, P., Liu, X.B., Jones, E.G., 2001. Differences in quantal amplitude reflect GluR4- subunit number at corticothalamic synapses on two populations of thalamic neurons. Proc Natl Acad Sci U S A 98(7), 4172-4177.

Grent-'t-Jong, T., Rivolta, D., Gross, J., Gajwani, R., Lawrie, S.M., Schwannauer, M., Heidegger, T., Wibral, M., Singer, W., Sauer, A., Scheller, B., Uhlhaas, P.J., 2018. Acute ketamine dysregulates task-related gamma-band oscillations in thalamo-cortical circuits in schizophrenia. Brain 141, 2511-2526.

Grunze, H.C., Rainnie, D.G., Hasselmo, M.E., Barkai, E., Hearn, E.F., McCarley, R.W., Greene, R.W., 1996. NMDA-dependent modulation of CA1 local circuit inhibition. J. Neurosci 16(6), 2034-2043.

Guillery, R.W., Sherman, S.M., 2002. Thalamic relay functions and their role in corticocortical communication: generalizations from the visual system. Neuron 33(2), 163-175.

Hakami, T., Jones, N.C., Tolmacheva, E.A., Gaudias, J., Chaumont, J., Salzberg, M., O'Brien, T.J., Pinault, D., 2009. NMDA receptor hypofunction leads to generalized and persistent aberrant gamma oscillations independent of hyperlocomotion and the state of consciousness. PLoS One 4(8), e6755.

Harris, R.M., Hendrickson, A.E., 1987. Local circuit neurons in the rat ventrobasal thalamus--a GABA immunocytochemical study. Neuroscience 21(1), 229-236.

Heimer, L., 2000. Basal forebrain in the context of schizophrenia. Brain Res Brain Res Rev 31(2-3), 205-235.

Hinze-Selch, D., Mullington, J., Orth, A., Lauer, C.J., Pollmacher, T., 1997. Effects of clozapine on sleep: a longitudinal study. Biol Psychiatry 42(4), 260-266.

Hobson, J.A., 1997. Dreaming as delirium: a mental status analysis of our nightly madness. Semin Neurol 17(2), 121-128.

Hoflich, A., Hahn, A., Kublbock, M., Kranz, G.S., Vanicek, T., Windischberger, C., Saria, A., Kasper, S., Winkler, D., Lanzenberger, R., 2015. Ketamine-Induced Modulation of the Thalamo-Cortical Network in Healthy Volunteers As a Model for Schizophrenia. Int. J Neuropsychopharmacol Apr 19;18(9). pii: pyv040. doi: 10.1093/ijnp/pyv040.

Homayoun, H., Moghaddam, B., 2007. NMDA receptor hypofunction produces opposite effects on prefrontal cortex interneurons and pyramidal neurons. J. Neurosci 27(43), 11496-11500.

Hong, L.E., Summerfelt, A., Buchanan, R.W., O'Donnell, P., Thaker, G.K., Weiler, M.A., Lahti, A.C., 2010. Gamma and delta neural oscillations and association with clinical symptoms under subanesthetic ketamine. Neuropsychopharmacology 35(3), 632-640.

Howland, R.H., 1997. Sleep-onset rapid eye movement periods in neuropsychiatric disorders: implications for the pathophysiology of psychosis. J Nerv Ment Dis 185(12), 730-738.

Huang, A.S., Rogers, B.P., Woodward, N.D., 2019. Disrupted modulation of thalamus activation and thalamocortical connectivity during dual task performance in schizophrenia. Schizophr Res 210, 270-277.

Hunt, M.J., Olszewski, M., Piasecka, J., Whittington, M.A., Kasicki, S., 2015. Effects of NMDA receptor antagonists and antipsychotics on high frequency oscillations recorded in the nucleus accumbens of freely moving mice. Psychopharmacology (Berl) 232(24), 4525-4535.

Jackson, M.E., Homayoun, H., Moghaddam, B., 2004. NMDA receptor hypofunction produces concomitant firing rate potentiation and burst activity reduction in the prefrontal cortex. Proc. Natl. Acad. Sci. U. S. A 101(22), 8467-8472.

Jacobsen, R.B., Ulrich, D., Huguenard, J.R., 2001. GABA(B) and NMDA receptors contribute to spindle-like oscillations in rat thalamus in vitro. J. Neurophysiol 86(3), 1365-1375.

Jones, N.C., Reddy, M., Anderson, P., Salzberg, M.R., O'Brien, T.J., Pinault, D., 2012. Acute administration of typical and atypical antipsychotics reduces EEG gamma power, but only the preclinical compound LY379268 reduces the ketamine-induced rise in gamma power. Int J Neuropsychopharmacol 15(5), 657-668.

Kamath, J., Virdi, S., Winokur, A., 2015. Sleep Disturbances in Schizophrenia. Psychiatr Clin North Am 38(4), 777-792.

Kambeitz, J., Kambeitz-Ilankovic, L., Cabral, C., Dwyer, D.B., Calhoun, V.D., van den Heuvel, M.P., Falkai, P., Koutsouleris, N., Malchow, B., 2016. Aberrant Functional Whole-Brain Network Architecture in Patients With Schizophrenia: A Meta-analysis. Schizophr Bull 42 Suppl 1, S13-21.

Kane, J.M., Honigfeld, G., Singer, J., Meltzer, H., 1988. Clozapine in treatment-resistant schizophrenics. Psychopharmacol Bull 24(1), 62-67.

Kaskie, R.E., Ferrarelli, F., 2018. Investigating the neurobiology of schizophrenia and other major psychiatric disorders with Transcranial Magnetic Stimulation. Schizophr Res 192, 30-38.

Kenny, J.D., Chemali, J.J., Cotten, J.F., Van Dort, C.J., Kim, S.E., Ba, D., Taylor, N.E., Brown, E.N., Solt, K., 2016. Physostigmine and Methylphenidate Induce Distinct Arousal States During Isoflurane General Anesthesia in Rats. Anesth Analg 123(5), 1210-1219.

Kim, T., Thankachan, S., McKenna, J.T., McNally, J.M., Yang, C., Choi, J.H., Chen, L., Kocsis, B., Deisseroth, K., Strecker, R.E., Basheer, R., Brown, R.E., McCarley, R.W., 2015. Cortically projecting basal forebrain parvalbumin neurons regulate cortical gamma band oscillations. Proc Natl Acad Sci U S A 112(11), 3535-3540.

Kocsis, B., 2012a. Differential role of NR2A and NR2B subunits in N-methyl-D-aspartate receptor antagonist-induced aberrant cortical gamma oscillations. Biol. Psychiatry 71(11), 987-995.

Kocsis, B., 2012b. State-dependent increase of cortical gamma activity during REM sleep after selective blockade of NR2B subunit containing NMDA receptors. Sleep 35(7), 1011-1016.

Krystal, J.H., Abdallah, C.G., Sanacora, G., Charney, D.S., Duman, R.S., 2019. Ketamine: A Paradigm Shift for Depression Research and Treatment. Neuron 101(5), 774-778.

Krystal, J.H., Karper, L.P., Seibyl, J.P., Freeman, G.K., Delaney, R., Bremner, J.D., Heninger, G.R., Bowers, M.B., Jr., Charney, D.S., 1994. Subanesthetic effects of the noncompetitive NMDA antagonist, ketamine, in humans. Psychotomimetic, perceptual, cognitive, and neuroendocrine responses. Arch. Gen. Psychiatry 51(3), 199-214.

Lam, R.W., Sherman, S.M., 2010. Functional organization of the somatosensory cortical layer 6 feedback to the thalamus. Cereb Cortex 20(1), 13-24.

Landisman, C.E., Connors, B.W., 2007. VPM and PoM nuclei of the rat somatosensory thalamus: intrinsic neuronal properties and corticothalamic feedback. Cereb Cortex 17(12), 2853-2865.

Lee, Y.J., Cho, S.J., Cho, I.H., Jang, J.H., Kim, S.J., 2012. The relationship between psychotic-like experiences and sleep disturbances in adolescents. Sleep Med 13(8), 1021-1027.

Lewicki, M.S., 1998. A review of methods for spike sorting: the detection and classification of neural action potentials. Network 9(4), R53-78.

Lipina, T., Labrie, V., Weiner, I., Roder, J., 2005. Modulators of the glycine site on NMDA receptors, D-serine and ALX 5407, display similar beneficial effects to clozapine in mouse models of schizophrenia. Psychopharmacology (Berl) 179(1), 54-67.

Lu, J., Nelson, L.E., Franks, N., Maze, M., Chamberlin, N.L., Saper, C.B., 2008. Role of endogenous sleep-wake and analgesic systems in anesthesia. J Comp Neurol 508(4), 648-662.

Manoach, D.S., Demanuele, C., Wamsley, E.J., Vangel, M., Montrose, D.M., Miewald, J., Kupfer, D., Buysse, D., Stickgold, R., Keshavan, M.S., 2014. Sleep spindle deficits in antipsychotic-naive early course schizophrenia and in non-psychotic first-degree relatives. Front Hum Neurosci 8, 762.

Manoach, D.S., Pan, J.Q., Purcell, S.M., Stickgold, R., 2016. Reduced Sleep Spindles in Schizophrenia: A Treatable Endophenotype That Links Risk Genes to Impaired Cognition? Biol Psychiatry 80(8), 599-608.

Manoach, D.S., Stickgold, R., 2019. Abnormal Sleep Spindles, Memory Consolidation, and Schizophrenia. Annu Rev Clin Psychol 15, 451-479.

Mason, O., Wakerley, D., 2012. The psychotomimetic nature of dreams: an experimental study. Schizophr Res Treatment 2012, 872307.

Moghaddam, B., Adams, B., Verma, A., Daly, D., 1997. Activation of glutamatergic neurotransmission by ketamine: a novel step in the pathway from NMDA receptor blockade to dopaminergic and cognitive disruptions associated with the prefrontal cortex. J. Neurosci 17(8), 2921-2927.

Monti, J.M., Monti, D., 2005. Sleep disturbance in schizophrenia. Int Rev Psychiatry 17(4), 247-253.

Moruzzi, G., Magoun, H.W., 1949. Brain stem reticular formation and activation of the EEG. Electroencephalogr. Clin. Neurophysiol 1(4), 455-473.

Mota, N.B., Resende, A., Mota-Rolim, S.A., Copelli, M., Ribeiro, S., 2016. Psychosis and the Control of Lucid Dreaming. Front Psychol 7, 294.

Mrzljak, L., Bergson, C., Pappy, M., Huff, R., Levenson, R., Goldman-Rakic, P.S., 1996. Localization of dopamine D4 receptors in GABAergic neurons of the primate brain. Nature 381(6579), 245-248.

Mulle, C., Madariaga, A., Deschênes, M., 1986. Morphology and electrophysiological properties of reticularis thalami neurons in cat: in vivo study of a thalamic pacemaker. J Neurosci 6(8), 2134-2145.

Munk, M.H., Roelfsema, P.R., Konig, P., Engel, A.K., Singer, W., 1996. Role of reticular activation in the modulation of intracortical synchronization. Science 272(5259), 271-274.

Nugent, A.C., Ballard, E.D., Gould, T.D., Park, L.T., Moaddel, R., Brutsche, N.E., Zarate, C.A., Jr., 2019. Ketamine has distinct electrophysiological and behavioral effects in depressed and healthy subjects. Mol Psychiatry 24(7), 1040-1052.

Paxinos, G., Watson, C., 1998. The rat brain in stereotaxic coordinates, Fourth edition ed. Academic Press.

Pinault, D., 1996. A novel single-cell staining procedure performed in vivo under electrophysiological control: morpho-functional features of juxtacellularly labeled thalamic cells and other central neurons with biocytin or Neurobiotin. J Neurosci Methods 65(2), 113-136.

Pinault, D., 2004. The thalamic reticular nucleus: structure, function and concept. Brain Res Brain Res Rev 46(1), 1-31.

Pinault, D., 2005. A new stabilizing craniotomy-duratomy technique for single-cell anatomo-electrophysiological exploration of living intact brain networks. J Neurosci Methods 141(2), 231-242.

Pinault, D., 2008. N-methyl d-aspartate receptor antagonists ketamine and MK-801 induce wake-related aberrant gamma oscillations in the rat neocortex. Biol Psychiatry 63(8), 730-735.

Pinault, D., 2011. Dysfunctional thalamus-related networks in schizophrenia. Schizophr Bull 37(2), 238-243.

Pinault, D., Deschenes, M., 1992. Muscarinic inhibition of reticular thalamic cells by basal forebrain neurones. Neuroreport 3(12), 1101-1104.

Pinault, D., Deschenes, M., 1998. Anatomical evidence for a mechanism of lateral inhibition in the rat thalamus. Eur J Neurosci 10(11), 3462-3469.

Pinault, D., Deschênes, M., 1992a. Control of 40-Hz firing of reticular thalamic cells by neurotransmitters. Neuroscience 51(2), 259-268.

Pinault, D., Deschênes, M., 1992b. Voltage-dependent 40-Hz oscillations in rat reticular thalamic neurons in vivo. Neuroscience 51(2), 245-258.

Pinault, D., Slezia, A., Acsady, L., 2006. Corticothalamic 5-9 Hz oscillations are more pro-epileptogenic than sleep spindles in rats. J Physiol 574(Pt 1), 209-227.

Pinault, D., Vergnes, M., Marescaux, C., 2001. Medium-voltage 5-9-Hz oscillations give rise to spike-and-wave discharges in a genetic model of absence epilepsy: in vivo dual extracellular recording of thalamic relay and reticular neurons. Neuroscience 105(1), 181-201.

Pitsikas, N., Boultadakis, A., Sakellaridis, N., 2008. Effects of sub-anesthetic doses of ketamine on rats' spatial and non-spatial recognition memory. Neuroscience 154(2), 454-460.

Poulin, J., Daoust, A.M., Forest, G., Stip, E., Godbout, R., 2003. Sleep architecture and its clinical correlates in first episode and neuroleptic-naive patients with schizophrenia. Schizophr Res 62(1-2), 147-153.

Pratt, J.A., Morris, B.J., 2015. The thalamic reticular nucleus: a functional hub for thalamocortical network dysfunction in schizophrenia and a target for drug discovery. J Psychopharmacol 29(2), 127-137.

Ramyead, A., Kometer, M., Studerus, E., Koranyi, S., Ittig, S., Gschwandtner, U., Fuhr, P., Riecher-Rossler, A., 2015. Aberrant Current Source-Density and Lagged Phase Synchronization of Neural Oscillations as Markers for Emerging Psychosis. Schizophr Bull 41(4), 919-929.

Rivolta, D., Heidegger, T., Scheller, B., Sauer, A., Schaum, M., Birkner, K., Singer, W., Wibral, M., Uhlhaas, P.J., 2015. Ketamine Dysregulates the Amplitude and Connectivity of High-Frequency Oscillations in Cortical-Subcortical Networks in Humans: Evidence From Resting-State Magnetoencephalography-Recordings. Schizophr. Bull 41(5), 1105-1114.

Roy, R.C., Stullken, E.H., 1981. Electroencephalographic evidence of arousal in dogs from halothane after doxapram, physostigmine, or naloxone. Anesthesiology 55(4), 392-397.

Scarone, S., Manzone, M.L., Gambini, O., Kantzas, I., Limosani, I., D'Agostino, A., Hobson, J.A., 2008. The dream as a model for psychosis: an experimental approach using bizarreness as a cognitive marker. Schizophr Bull 34(3), 515-522.

Schwieler, L., Linderholm, K.R., Nilsson-Todd, L.K., Erhardt, S., Engberg, G., 2008. Clozapine interacts with the glycine site of the NMDA receptor: electrophysiological studies of dopamine neurons in the rat ventral tegmental area. Life Sci 83(5-6), 170-175.

Sherman, S.M., Koch, C., 1986. The control of retinogeniculate transmission in the mammalian lateral geniculate nucleus. Exp Brain Res 63(1), 1-20.

Sitaram, N., Wyatt, R.J., Dawson, S., Gillin, J.C., 1976. REM sleep induction by physostigmine infusion during sleep. Science 191(4233), 1281-1283.

Sleigh, J., Harvey, M., Voss, L., denny, B., 2014. Ketamine - More mechanisms of action than just NMDA blockade. Trends in Anaesthesia and Critical Care 4, 76-81.

Slovik, M., Rosin, B., Moshel, S., Mitelman, R., Schechtman, E., Eitan, R., Raz, A., Bergman, H., 2017. Ketamine induced converged synchronous gamma oscillations in the cortico-basal ganglia network of nonhuman primates. J Neurophysiol 118(2), 917-931.

Snyder, M.A., Gao, W.J., 2019. NMDA receptor hypofunction for schizophrenia revisited: Perspectives from epigenetic mechanisms. Schizophr Res. Apr 9. pii: S0920-9964(19)30104-5. doi: 10.1016/j.schres.**2019**.03.010. [Epub ahead of print]

Spencer, K.M., Nestor, P.G., Perlmutter, R., Niznikiewicz, M.A., Klump, M.C., Frumin, M., Shenton, M.E., McCarley, R.W., 2004. Neural synchrony indexes disordered perception and cognition in schizophrenia. Proc. Natl. Acad. Sci. U. S. A 101(49), 17288-17293.

Steriade, M., Deschênes, M., Domich, L., Mulle, C., 1985. Abolition of spindle oscillations in thalamic neurons disconnected from nucleus reticularis thalami. J Neurophysiol 54(6), 1473-1497.

Steriade, M., McCormick, D.A., Sejnowski, T.J., 1993. Thalamocortical oscillations in the sleeping and aroused brain. Science 262(5134), 679-685.

Steullet, P., 2019. Thalamus-related anomalies as candidate mechanism-based biomarkers for psychosis. Schizophr Res. May 27. pii: S0920-9964(19)30203-8. doi: 10.1016/j.schres.**2019**.05.027. [Epub ahead of print]

Tieges, Z., McGrath, A., Hall, R.J., Maclullich, A.M., 2013. Abnormal level of arousal as a predictor of delirium and inattention: an exploratory study. Am J Geriatr Psychiatry 21(12), 1244-1253.

Tsekou, H., Angelopoulos, E., Paparrigopoulos, T., Golemati, S., Soldatos, C.R., Papadimitriou, G.N., Ktonas, P.Y., 2015. Sleep EEG and spindle characteristics after combination treatment with clozapine in drug-resistant schizophrenia: a pilot study. J Clin Neurophysiol 32(2), 159-163.

Vukadinovic, Z., 2014. NMDA receptor hypofunction and the thalamus in schizophrenia. Physiol Behav 131, 156-159.

Wamsley, E.J., Tucker, M.A., Shinn, A.K., Ono, K.E., McKinley, S.K., Ely, A.V., Goff, D.C., Stickgold, R., Manoach, D.S., 2012. Reduced sleep spindles and spindle coherence in schizophrenia: mechanisms of impaired memory consolidation? Biol Psychiatry 71(2), 154-161.

Woodward, N.D., Heckers, S., 2016. Mapping Thalamocortical Functional Connectivity in Chronic and Early Stages of Psychotic Disorders. Biol Psychiatry 79(12), 1016-1025.

Young, A., Wimmer, R.D., 2016. Implications for the thalamic reticular nucleus in impaired attention and sleep in schizophrenia. Schizophr Res 180, 44-47.

Zanini, M.A., Castro, J., Cunha, G.R., Asevedo, E., Pan, P.M., Bittencourt, L., Coelho, F.M., Tufik, S., Gadelha, A., Bressan, R.A., Brietzke, E., 2015. Abnormalities in sleep patterns in individuals at risk for psychosis and bipolar disorder. Schizophr Res 169(1-3), 262-267.

Zanos, P., Moaddel, R., Morris, P.J., Riggs, L.M., Highland, J.N., Georgiou, P., Pereira, E.F.R., Albuquerque, E.X., Thomas, C.J., Zarate, C.A., Jr., Gould, T.D., 2018. Ketamine and Ketamine Metabolite Pharmacology: Insights into Therapeutic Mechanisms. Pharmacol Rev 70(3), 621-660.